\documentclass[12pt]{iopart}
\usepackage[utf8]{inputenc}
\usepackage[T1]{fontenc}
\usepackage{xspace}
\expandafter\let\csname equation*\endcsname=\relax
\expandafter\let\csname endequation*\endcsname=\relax
\usepackage{amsmath,amssymb,amsfonts,amsthm}

\theoremstyle{definition}

\theoremstyle{remark}

\catcode`,\active

\catcode`\,12

\newcommand*\pFqskip{8mu}
\catcode`,\active
\newcommand*\pFq{\begingroup
        \catcode`\,\active
        \def ,{\mskip\pFqskip\relax}%
        \dopFq
}
\catcode`\,12
\def\dopFq#1#2#3#4#5{%
        {}_{#1}F_{#2}\biggl[\genfrac..{0pt}{}{#3}{#4};#5\biggr]%
        \endgroup
}

\renewcommand{\phi}{\varphi}
\newcommand{\pd}{\partial}
\newcommand{\ket}[1]{\ensuremath{|#1\rangle}\xspace}

\newcommand{\under}[1]{_{#1}}

\begin{document}
\title[The Dunkl oscillator in the plane II]{The Dunkl oscillator in the plane II : representations of the symmetry algebra}
\author{Vincent X. Genest}
\ead{genestvi@crm.umontreal.ca}
\address{Centre de recherches math\'ematiques, Universit\'e de Montr\'eal, C.P. 6128, Succursale Centre-ville, Montr\'eal, Qu\'ebec, Canada, H3C 3J7}
\author{Mourad E.H. Ismail}
\ead{mourad.eh.ismail@gmail.com}
\address{Department of Mathematics, University of Central Florida, Orlando, FL 32816, USA}
\address{Department of Mathematics, King Saud University, Riyadh, Saudi Arabia}
\author{Luc Vinet}
\ead{luc.vinet@umontreal.ca}
\address{Centre de recherches math\'ematiques, Universit\'e de Montr\'eal, C.P. 6128, Succursale Centre-ville, Montr\'eal, Qu\'ebec, Canada, H3C 3J7}
\author{Alexei Zhedanov}
\ead{zhedanov@yahoo.com}
\address{Donetsk Institute for Physics and Technology, Donetsk 83114, Ukraine}
\begin{abstract}
The superintegrability, wavefunctions and overlap coefficients of the Dunkl oscillator model in the plane were considered in the first part. Here finite-dimensional representations of the symmetry algebra of the system, called the Schwinger-Dunkl algebra $sd(2)$, are investigated. The algebra $sd(2)$ has six generators, including two involutions and a central element, and can be seen as a deformation of the Lie algebra $\mathfrak{u}(2)$. Two of the symmetry generators, $J_3$ and $J_2$, are respectively associated to the separation of variables in Cartesian and polar coordinates. Using the parabosonic creation/annihilation operators, two bases for the representations of $sd(2)$, the Cartesian and circular bases, are constructed. In the Cartesian basis, the operator $J_3$ is diagonal and the operator $J_2$ acts in a tridiagonal fashion. In the circular basis, the operator $J_2$ is block upper-triangular with all blocks $2\times 2$ and the operator $J_3$ acts in a tridiagonal fashion. The expansion coefficients between the two bases are given by the Krawtchouk polynomials. In the general case, the eigenvectors of $J_2$ in the circular basis are generated by the Heun polynomials and their components are expressed in terms of the para-Krawtchouk polynomials. In the fully isotropic case, the eigenvectors of $J_2$ are generated by little $-1$ Jacobi or ordinary Jacobi polynomials. The basis in which the operator $J_2$ is diagonal is then considered. In this basis, the defining relations of the Schwinger-Dunkl algebra imply that $J_3$ acts in a block tridiagonal fashion with all blocks $2\times 2$. The matrix elements of $J_3$ in this basis are given explicitly.
\end{abstract}
\section{Introduction}
This is the second part of this series concerned with the analysis of the isotropic Dunkl oscillator model. In part I, the model has been shown to be superintegrable, the wavefunctions have been obtained in Cartesian and polar coordinates and the overlap coefficients have been found \cite{Genest-2012-4}. In the present work, the representations of the symmetry algebra of the model, called the Schwinger-Dunkl algebra (see below), are investigated. As shall be seen, this study entails remarkable connections with special functions such as the Heun, little $-1$ Jacobi and para-Krawtchouk polynomials.
\subsection{Superintegrability}
One recalls that a quantum system defined by a Hamiltonian $H$ in $d$ dimensions is maximally \emph{superintegrable} if it admits $2d-1$ algebraically independent symmetry generators $S_i$ that commute with the Hamiltonian
\begin{align*}
[S_i,H]=0,\qquad 1\leqslant i\leqslant 2d-1,
\end{align*}
where one of the symmetries is the Hamiltonian itself, e.g. $S_1\equiv H$. Moreover, a superintegrable system is said to be of order $\ell$ if $\ell$ is the maximal order of the symmetries $S_i$ in the momentum variables.
\subsection{The Dunkl oscillator model}
The isotropic Dunkl oscillator model \cite{DeBie-2011,Genest-2012-4,Nowak-2009} in the plane is possibly the simplest two-dimensional system described by a Hamiltonian involving reflections. It is second-order superintegrable and is defined by the Hamiltonian \cite{Genest-2012-4}
\begin{equation}
\label{Hamiltonian}
\mathcal{H}=-\frac{1}{2}\left[(\mathcal{D}_{x}^{\mu_{x}})^2+(\mathcal{D}_{y}^{\mu_{y}})^2\right]+\frac{1}{2}[x^2+y^2],
\end{equation}
where $\mathcal{D}_{x_i}^{\mu_{x_i}}$ is the Dunkl derivative \cite{Dunkl-2002,Rozen-1994}
\begin{align*}
\mathcal{D}_{x_i}^{\mu_{x_i}}=\pd_{x_{i}}+\frac{\mu_{x_i}}{x_{i}}(\mathbb{I}-R_{x_{i}}),\qquad \pd_{x_i}=\frac{\pd}{\pd x_i},
\end{align*}
with $\mathbb{I}$ denoting the identity operator and $R_{x_i}$, $x_i\in\{x,y\}$, standing for the reflection operator with respect to the plane $x_i=0$. Hence the reflections $R_{x}$, $R_{y}$ that appear in the Hamiltonian \eqref{Hamiltonian} have the action
\begin{align*}
R_x f(x,y)=f(-x,y),\qquad R_yf(x,y)=f(x,-y),
\end{align*}
and thus evidently $R_{x_i}^2=\mathbb{I}$. In connection with the nomenclature of the standard harmonic oscillator, the model is called isotropic because the quadratic potential is $SO(2)$ invariant. For the full Hamiltonian \eqref{Hamiltonian} to have this symmetry requires of course that $\mu_x=\mu_y$. 

The Schr\"odinger equation associated to $\mathcal{H}$ is separable in both Cartesian and polar coordinates. The spectrum of energies $\mathcal{E}$ is given by
\begin{align}
\label{Energy}
\mathcal{E}_{N}=N+\mu_x+\mu_y+1,\qquad N=n_x+n_y,
\end{align}
where $n_x$, $n_y$ are non-negative integers. The wavefunctions are well defined for the values $\mu_x,\mu_y\in (-\frac{1}{2},\infty)$; the case $\mu_x=\mu_y=0$ corresponds to the standard quantum harmonic oscillator. It is easily seen from \eqref{Energy} that the energy level $\mathcal{E}_{N}$ exhibits a $N+1$-fold degeneracy.
\subsection{Symmetries of the Dunkl oscillator}
The symmetries of the Dunkl oscillator Hamiltonian \eqref{Hamiltonian} can be obtained by the Schwinger construction using the parabosonic creation/annihilation operators \cite{Genest-2012-4,Green-1953,Mukunda-1980}. We  consider the operators \cite{Rozen-1994,Zhedanov-2011}
\begin{align}
\label{Creation}
A_{\pm}^{x_i}=\frac{1}{\sqrt{2}}(x_i\pm\mathcal{D}_{x_i}^{\mu_{x_i}}),\qquad x_i\in\{x,y\}.
\end{align}
It is verified that the operators $A^{x_i}_{\pm}$ satisfy the following commutation relations:
\begin{gather}
\label{Commutation}
[A_{-}^{x_i},A_{+}^{x_i}]=\mathbb{I}+2\mu_{x_i}R_{x_i},\qquad \{A_{\pm}^{x_i},R_{x_i}\}=0,
\end{gather}
where $\{x,y\}=xy+yx$ denotes the anticommutator. In addition to the commutation relations \eqref{Commutation}, one has
\begin{align}
\label{Additional}
[A_{\pm}^{x_i},A_{\pm}^{x_j}]=[A_{\pm}^{x_i},R_{x_j}]=[R_{x_i},R_{x_j}]=0,\qquad i\neq j.
\end{align}
In terms of the operators \eqref{Creation}, the Hamiltonian \eqref{Hamiltonian} takes the form
\begin{align*}
\mathcal{H}=\frac{1}{2}\{A^{x}_{+},A^{x}_{-}\}+\frac{1}{2}\{A^{y}_{+},A^{y}_{-}\}=\mathcal{H}_x+\mathcal{H}_y,
\end{align*}
where
\begin{align}
\label{Voila-2}
\mathcal{H}_{x_i}=\frac{1}{2}\{A_{+}^{x_i},A_{-}^{x_i}\}=-\frac{1}{2}(\mathcal{D}^{\mu_{x_i}}_{x_i})^2+\frac{1}{2}x_i^2,
\end{align}
is the Hamiltonian of the one-dimensional Dunkl oscillator.

The symmetry generators of the Dunkl oscillator model are as follows. Consider the operator
\begin{align}
\label{Realization-3}
J_3=\frac{1}{4}\{A^{x}_{-},A^{x}_{+}\}-\frac{1}{4}\{A^{y}_{-},A^{y}_{+}\}.
\end{align}
It is directly verified that $[\mathcal{H},J_3]=0$. Since $J_3$ can be written as
\begin{align*}
J_3=\frac{1}{2}(\mathcal{H}_x-\mathcal{H}_y),
\end{align*}
using \eqref{Voila-2}, it is clear that this symmetry corresponds to the separability of the Schr\"odinger equation in Cartesian coordinates \cite{Genest-2012-4}. A second symmetry generator is given by
\begin{align}
\label{Realization-2}
J_2=\frac{1}{2i}(A^{x}_{+}A^{y}_{-}-A^{x}_{-}A^{y}_{+}).
\end{align}
It is again directly verified that $[J_2,\mathcal{H}]=0$. In terms of Dunkl derivatives, this operator has the expression
\begin{align*}
J_2=\frac{1}{2i}\left(x\mathcal{D}_{y}^{\mu_{y}}-y\mathcal{D}_{x}^{\mu_{x}}\right),
\end{align*}
and it has been shown \cite{Genest-2012-4} that $J_2$ is the symmetry corresponding to the separation of variables in polar coordinates. A third symmetry $J_1$, algebraically dependent of $J_2$, $J_3$, is obtained by taking $J_1=-i[J_2,J_3]$. This additional symmetry generator reads
\begin{align}
\label{Realization-1}
J_1=\frac{1}{2}(A^{x}_{+}A^{y}_{-}+A^{x}_{-}A^{y}_{+}).
\end{align}
In addition to $J_i$, $i=1,\ldots,3$, it is directly checked that the reflections $R_{x}$, $R_y$ also commute with $\mathcal{H}$.
\subsection{The main object: the Schwinger-Dunkl algebra $sd(2)$}
The symmetry algebra of the Dunkl oscillator, called the Schwinger-Dunkl algebra, is denoted $sd(2)$ and defined by the commutation relations
\begin{subequations}
\label{SD-2}
\begin{gather}
\{J_1,R_{x_i}\}=0,\quad \{J_2,R_{x_i}\}=0,\quad [J_3,R_{x_i}]=0,\\
[J_2,J_3]=iJ_1,\qquad [J_3,J_1]=iJ_2,\\
[J_1,J_2]=i\Big(J_3+J_3(\mu_x R_{x}+\mu_y R_y)-\mathcal{H}(\mu_xR_x-\mu_y R_y)/2\Big),
\end{gather}
\end{subequations}
where $R_{x_i}^2=\mathbb{I}$, $x_i\in\{x,y\}$, and the Hamiltonian $\mathcal{H}$ is a central element.  The algebra $sd(2)$ admits the Casimir operator \cite{Genest-2012-2}
\begin{align*}
C=J_1^2+J_2^2+J_3^2+\mu_x R_x/2+\mu_yR_y/2+\mu_x\mu_yR_xR_y,
\end{align*}
which commutes with all the generators. In the present realization, the Casimir operator $C$ takes the value
\begin{align*}
C=\frac{1}{4}(\mathcal{H}^2-1).
\end{align*}
Note that the involution $P=R_xR_y$ also commutes with all the generators and thus can be viewed as a second Casimir operator. Furthermore, it is easily seen that when $\mu_x=\mu_y=0$, the Schwinger-Dunkl algebra $sd(2)$ reduces to the Lie algebra $\mathfrak{u}(2)$, which is the symmetry algebra of the standard isotropic 2D quantum oscillator in the plane.

The irreducible representations of $sd(2)$ can be used to account for the degeneracies in the spectrum of $\mathcal{H}$. In finite-dimensional representations of degree $N+1$, the action of the symmetry generators $J_1$, $J_2$, $J_3$, $R_x$ and $R_y$ indicate how the degenerate eigenstates of $\mathcal{H}$ corresponding to the energy value $\mathcal{E}_{N}$ transform into one another under the action of the symmetries. In the following, three bases for the finite-dimensional irreducible representations of $sd(2)$ will be constructed and the explicit formulas for the action of the symmetry generators on each basis will be derived.
\subsection{Outline}
Here is the outline of the paper. In Section 2, we construct the Cartesian basis in which the symmetry generator $J_3$ is diagonal and $J_2$ acts in a tridiagonal fashion. In Section 3, we introduce the circular creation/annihilation operators and study the associated circular basis in which $J_2$ is block upper-triangular and $J_3$ is tridiagonal. We show that the interbasis expansion coefficients involve the Krawtchouk polynomials and we derive the spectrum of $J_2$ algebraically. In Section 4, we obtain the eigenvectors of $J_2$ in the circular basis for odd-dimensional representations and show that these eigenvectors are generated by the Heun polynomials and that their components are para-Krawtchouk polynomials. The fully isotropic case is shown to involve the little $-1$ Jacobi polynomials. In Section 5, the eigenvectors of $J_2$ in the circular basis for even-dimensional representations are studied. In Section 6, we examine the basis in which $J_2$ is diagonal and show that $J_3$ acts in a six-diagonal fashion on this basis. We conclude with an outlook.
\section{The Cartesian basis}
In this section the Cartesian basis for the finite-dimensional representations of $sd(2)$ is constructed using the realization \eqref{Realization-3}, \eqref{Realization-2}, \eqref{Realization-1} of the algebra generators in terms of the creation/annihilation operators \eqref{Creation}. The representation spaces spanned by the Cartesian basis correspond to the spaces of degenerate wavefunctions with energies $\mathcal{E}_{N}$,$N\in\mathbb{N}$, separated in Cartesian coordinates, although a different normalization is used for the basis vectors. The action of the $sd(2)$ generators on the wavefunctions were obtained by a direct computation in \cite{Genest-2012-4} using the expressions of the symmetries in terms of Dunkl derivatives. Here the actions of the generators and the spectra of the Hamiltonian $\mathcal{H}$ and the symmetry generator $J_3$ are obtained in a purely algebraic manner.

The Cartesian basis vectors are labeled by two non-negative integers $n_x$, $n_y$ and are denoted by $\ket{n_x,n_y}$. These basis vectors are defined by
\begin{align}
\label{Def-Cart}
\ket{n_x,n_y}=(A_{+}^{x})^{n_x}(A_{+}^{y})^{n_y}\ket{0_x,0_y},
\end{align}
where $\ket{0_x,0_y}$ is the ''vacuum'' vector. The vacuum vector has the defining properties
\begin{subequations}
\begin{gather}
\label{Vacuum-Anni}
A_{-}^{x}\ket{0_x,0_y}=0,\quad A_{-}^{y}\ket{0_x,0_y}=0,\\
\label{Vacuum-Parity}
R_{x}\ket{0_x,0_y}=\ket{0_x,0_y},\quad R_{y}\ket{0_x,0_y}=\ket{0_x,0_y}.
\end{gather}
\end{subequations}
The action of the reflection operators and the creation/annihilation operators on the Cartesian basis vectors can be derived from the above definitions and the commutation/anticommutation relations \eqref{Commutation} and \eqref{Additional}. From the anticommutation relations
\begin{align*}
\{A_{+}^{x},R_{x}\}=0,\quad \{A_{+}^{y},R_{y}\}=0,
\end{align*}
and the vacuum parity conditions \eqref{Vacuum-Parity}, it directly follows that
\begin{align}
\label{R}
R_{x}\ket{n_x,n_y}=(-1)^{n_x}\ket{n_x,n_y},\quad R_y\ket{n_x,n_y}=(-1)^{n_y}\ket{n_x,n_y}.
\end{align}
By the definition of the basis vectors \eqref{Def-Cart}, one has also
\begin{align}
\label{A-Plus}
A_{+}^{x}\ket{n_x,n_y}=\ket{n_x+1,n_y},\quad A_{+}^{y}\ket{n_x,n_y}=\ket{n_x,n_y+1}.
\end{align}
To derive the action of the operators $A_{-}^{x_i}$ on the Cartesian basis, one needs the commutator identity
\begin{align}
\label{Com-Iden-1}
[A_{-}^{x_i},(A_{+}^{x_i})^{n}]=(A_{+}^{x_i})^{n-1}\left[n+\mu_{x_i}(1-(-1)^{n})R_{x_i}\right],
\end{align}
which is easily proven by induction. It is convenient to introduce the $\mu$-numbers \cite{Rozen-1994}
\begin{align}
\label{Mu-n}
[n]_{\mu}=n+\mu(1-(-1)^{n}).
\end{align}
Using the identity \eqref{Com-Iden-1} and the formulas \eqref{Def-Cart}, \eqref{Vacuum-Anni} and \eqref{Vacuum-Parity}, one finds
\begin{align}
\label{A-Minus}
A_{-}^{x}\ket{n_x,n_y}=(A_{+}^{y})^{n_{y}}[A_{-}^{x},(A_{+}^{x})^{n_x}]\ket{0_x,0_y}=[n_x]_{\mu_x}\ket{n_x-1,n_y},
\end{align}
and similarly for $A_{-}^{y}$.

Using the results \eqref{R}, \eqref{A-Plus} and \eqref{A-Minus} along with the expressions of the symmetry generators $J_{i}$, $i=1,2,3$, and $\mathcal{H}$, in terms of the operators $A_{\pm}^{x_i}$ given in \eqref{Realization-3}, \eqref{Realization-2} and \eqref{Realization-1}, one finds that the action of the symmetries on the Cartesian basis is given by
\begin{subequations}
\begin{gather}
J_2\ket{n_x,n_y}=\frac{1}{2i}\Big([n_y]_{\mu_y}\ket{n_x+1,n_y-1}-[n_{x}]_{\mu_{x}}\ket{n_x-1,n_y+1}\Big),\\
J_3\ket{n_x,n_y}=\frac{1}{2}\left(n_x-n_y+\mu_x-\mu_y\right)\ket{n_x,n_y},
\end{gather}
\end{subequations}
and the action of $J_1$ can be obtained directly by commuting $J_2$ and $J_3$. 
The central element $\mathcal{H}$ has the action
\begin{align*}
\mathcal{H}\ket{n_x,n_y}=(n_x+n_y+\mu_x+\mu_y+1)\ket{n_x,n_y}.
\end{align*}
Hence the spectra of the symmetry generator $J_3$ and of the full Hamiltonian $\mathcal{H}$ of the Dunkl oscillator have been recovered in a purely algebraic manner. As is expected, the symmetry operators $J_1,\ldots J_3$ and the involutions $R_x$, $R_y$ transform the set of vectors $\ket{n_x,n_y}$ with a given value of $N=n_x+n_y$ into one another; these vectors are the degenerate eigenvectors of $\mathcal{H}$ with energy $\mathcal{E}_{N}=N+\mu_x+\mu_y+1$. 

The preceding results can be used to define an infinite family of $N+1$-dimensional irreducible modules of the Schwinger-Dunkl algebra $sd(2)$ \eqref{SD-2}. Let $\mu_x,\,\mu_y\in \mathbb{R}$ be real numbers such that $\mu_x,\,\mu_y\in(-1/2,\infty)$ and denote by $V^{(\mu_x,\mu_y)}_{N}$ the $N+1$-dimensional $\mathbb{C}$-vector space spanned by the basis vectors $v_{n}^{(\mu_x,\mu_y)}$, $n\in\{0,\ldots,N\}$. Consider the vector space $V_{N}^{(\mu_x,\mu_y)}$ endowed with the following actions of the $sd(2)$ generators:
\begin{subequations}
\begin{gather}
J_1v_{n}^{(\mu_x,\mu_y)}=\frac{1}{2}\left([N-n]_{\mu_y}v_{n+1}^{(\mu_x,\mu_y)}+[n]_{\mu_x}v_{n-1}^{(\mu_x,\mu_y)}\right),\\
J_2v_{n}^{(\mu_x,\mu_y)}=\frac{1}{2i}\left([N-n]_{\mu_y}v_{n+1}^{(\mu_x,\mu_y)}-[n]_{\mu_x}v_{n-1}^{(\mu_x,\mu_y)}\right),\\
\label{J3-Cartesian}
J_3v_{n}^{(\mu_x,\mu_y)}=\left(n+\frac{1}{2}(\mu_x-\mu_y-N)\right)v_{n}^{(\mu_x,\mu_y)},\\
R_xv_{n}^{(\mu_x,\mu_y)}=(-1)^{n}v_{n}^{(\mu_x,\mu_y)},\qquad R_{y}v_{n}^{(\mu_x,\mu_y)}=(-1)^{N-n}v_{n}^{(\mu_x,\mu_y)},
\end{gather}
\end{subequations}
where $[n]$ denotes the $\mu$-numbers \eqref{Mu-n}. The central element $\mathcal{H}$ and the Casimir operator have the actions
\begin{align*}
\mathcal{H}v_{n}^{(\mu_x,\mu_y)}=(N+\mu_x+\mu_y+1) v_{n}^{(\mu_x,\mu_y)},
\end{align*}
and 
\begin{align*}
Cv_{n}^{(\mu_x,\mu_y)}=\frac{1}{4}\left\{(N+\mu_x+\mu_y)(N+2+\mu_x+\mu_y)\right\}v_{n}^{(\mu_x,\mu_y)}.
\end{align*}
It is clear that $V_{N}^{(\mu_x,\mu_y)}$ is a $sd(2)$-module and its irreducibility follows from the fact that the $\mu$-numbers appearing in the matrix elements of $J_1$, $J_2$ are never zero for $\mu_x,\mu_y\in(-1/2,\infty)$. For $\mu_x=\mu_y=0$, it is directly seen that the $sd(2)$-module $V_{N}^{(0,0)}$ reduces to the standard $N+1$-dimensional irreducible $\mathfrak{su}(2)$ module.

\section{The circular basis}
In this section, the circular basis for the finite-dimensional representations of $sd(2)$ is constructed using the left/right circular operators. The actions of the symmetries on this basis are obtained and the spectrum of the generator $J_2$ is derived from these actions. The expansion coefficients between the circular and Cartesian bases, which involve the Krawtchouk polynomials, are also examined.

The left/right circular operators for the 2D Dunkl oscillator are introduced following the analogous construction in the standard 2D harmonic oscillator \cite{Cohen-1997}. We define
\begin{align}
\label{Left-Right}
A_{\pm}^{L}=\frac{1}{\sqrt{2}}\left(A^{x}_{\pm}\mp i A^{y}_{\pm}\right),\quad A_{\pm}^{R}=\frac{1}{\sqrt{2}}\left(A^{x}_{\pm}\pm i A^{y}_{\pm}\right),
\end{align}
where $A_{\pm}^{x_i}$ are the creation/annihilation operators of the Dunkl oscillator that obey the commutation relations \eqref{Commutation}. The inverse relations are easily seen to be
\begin{align*}
A^{x}_{\pm}=\frac{1}{\sqrt{2}}\left(A^{L}_{\pm}+A^{R}_{\pm}\right),\quad A^{y}_{\pm}=\frac{\pm i}{\sqrt{2}}\left(A^{L}_{\pm}-A^{R}_{\pm}\right).
\end{align*}
The left/right operators obey the commutation relations
\begin{gather*}
[A^{L}_{-},A^{R}_{-}]=0,\quad [A^{L}_{+},A^{R}_{+}]=0,\\
[A^{R}_{-},A^{L}_{+}]=\mu_xR_x-\mu_y R_y,\quad [A^{L}_{-},A^{R}_{+}]=\mu_x R_x-\mu_y R_y,\\
[A^{L}_{-},A^{L}_{+}]=\mathbb{I}+\mu_x R_x+\mu_y R_y,\quad [A^{R}_{-},A^{R}_{+}]=\mathbb{I}+\mu_x R_x+\mu_y R_y,
\end{gather*}
and the algebraic relations involving the reflections become
\begin{gather}
\label{Circular-Reflections}
R_xA^{L}_{\pm}=-A^{R}_{\pm}R_{x},\quad R_{x}A^{R}_{\pm}=-A^{L}_{\pm}R_{x},\quad
R_{y}A^{L}_{\pm}=A^{R}_{\pm}R_y,\quad R_{y}A^{R}_{\pm}=A^{L}_{\pm}R_y.
\end{gather}
The circular basis vectors $\ket{n_{L},n_{R}}$ are labeled by the two non-negative integers $n_{L}$, $n_{R}$ and are defined by
\begin{align}
\label{Def-Circular}
\ket{n_{L},n_{R}}=(A^{L}_{+})^{n_{L}}(A^{R}_{+})^{n_{R}}\ket{0_L,0_R},
\end{align}
where $\ket{0_L,0_R}$ is the circular vacuum vector with the properties
\begin{subequations}
\begin{gather}
\label{Vacuum-Circular}
A_{-}^{L}\ket{0_L,0_R}=0,\quad A_{-}^{R}\ket{0_L,0_R}=0,\\
\label{Parity-Vacuum-Circular}
R_x\ket{0_L,0_R}=\ket{0_L,0_R},\quad R_y\ket{0_L,0_R}=\ket{0_L,0_R}.
\end{gather}
\end{subequations}
Given the definition \eqref{Def-Circular}, one has 
\begin{align*}
A_{+}^{L}\ket{n_L,n_R}=\ket{n_L+1,n_R},\quad A_{+}^{R}\ket{n_L,n_R}=\ket{n_L,n_R+1}.
\end{align*}
From the relations \eqref{Circular-Reflections} and the definition \eqref{Def-Circular}, it follows that
\begin{align}
\label{Action-Reflection}
R_{x}\ket{n_L,n_R}=(-1)^{n_L+n_R}\ket{n_R,n_L},\quad R_{y}\ket{n_L,n_R}=\ket{n_R,n_L}.
\end{align}
Consider the commutator identities
\begin{gather*}
[A^{L}_{-},(A^{L}_{+})^{n+1}]=(n+1)(A^{L}_{+})^{n}+\sum_{\alpha=0}^{n}(A^{L}_{+})^{n-\alpha}(A^{R}_{+})^{\alpha}\left\{(-1)^{\alpha}\mu_xR_x+\mu_yR_y\right\},\\
[A^{L}_{-},(A_{+}^{R})^{n+1}]=\sum_{\beta=0}^{n}(A^{L}_{+})^{n-\beta}(A^{R}_{+})^{\beta}\left\{(-1)^{n-\beta}\mu_xR_x-\mu_yR_y\right\},
\end{gather*}
which can be proven straightforwardly by induction. From the definition \eqref{Def-Circular}, the vacuum conditions \eqref{Vacuum-Circular}, \eqref{Parity-Vacuum-Circular} and the above identities, the action of $A_{-}^{L}$ on the circular basis elements $\ket{n_L,n_r}$ can be derived by a direct computation. For $n_L=n_R$, one has
\begin{gather*}
A_{-}^{L}\ket{n_L,n_R}=n_L\ket{n_L-1,n_R},
\end{gather*}
For $n_L>n_R$, one finds
\begin{gather*}
A_{-}^{L}\ket{n_L,n_R}=n_L\ket{n_L-1,n_R}+\sum_{j=n_R}^{n_L-1}\{(-1)^{n_R+j}\mu_x+\mu_y\}\ket{n_L+n_R-j-1,j}.
\end{gather*}
Finally, for $n_L<n_R$, one obtains
\begin{gather*}
A_{-}^{L}\ket{n_L,n_R}=n_L\ket{n_L-1,n_R}-\sum_{j=n_L}^{n_R-1}\{(-1)^{n_R+j}\mu_x+\mu_y\}\ket{n_L+n_R-j-1,j}.
\end{gather*}
To obtain the corresponding formulas for the action of $A^{R}_{-}$, one needs the identities
\begin{gather*}
[A^{R}_{-},(A^{R}_{+})^{n+1}]=(n+1)(A^{R}_{+})^{n}+\sum_{\alpha=0}^{n}(A^{L}_{+})^{n-\alpha}(A^{R}_{+})^{\alpha}\left\{(-1)^{n-\alpha}\mu_xR_x+\mu_yR_y\right\},\\
[A^{R}_{-},(A_{+}^{L})^{n+1}]=\sum_{\beta=0}^{n}(A^{L}_{+})^{n-\beta}(A^{R}_{+})^{\beta}\left\{(-1)^{\beta}\mu_xR_x-\mu_yR_y\right\}.
\end{gather*}
Using the same procedure as for $A_{-}^{L}$, we obtain the action of $A_{-}^{R}$. For $n_L=n_R$, we have
\begin{gather*}
A_{-}^{R}\ket{n_L,n_R}=n_R\ket{n_L,n_R-1}.
\end{gather*}
When $n_L>n_R$, one finds
\begin{gather*}
A_{-}^{R}\ket{n_L,n_R}=n_R\ket{n_L,n_R-1}+\sum_{j=n_R}^{n_L-1}\{(-1)^{n_R+j}\mu_x-\mu_y\}\ket{n_L+n_R-j-1,j},
\end{gather*}
and for $n_R>n_L$,  the result is
\begin{gather*}
A_{-}^{R}\ket{n_L,n_R}=n_R\ket{n_L,n_R-1}-\sum_{j=n_L}^{n_R-1}\{(-1)^{n_R+j}\mu_x-\mu_y\}\ket{n_L+n_R-j-1,j}.
\end{gather*}
As is seen from the formulas, the operators $A_{-}^{L/R}$ have the effect of sending the circular basis vectors $\ket{n_L,n_R}$ to \emph{all}  circular basis vectors $\ket{i_Lj_R}$ with $i_L+j_R=n_L+n_R-1$.

In terms of the circular operators \eqref{Left-Right}, the symmetry generators and the central element $\mathcal{H}$ take the rather symmetric form
\begin{subequations}
\begin{gather}
\label{Circular-Symmetries-1}
J_1=\frac{i}{4}\Big(\{A^{L}_{+},A^{R}_{-}\}-\{A^{L}_{-},A^{R}_{+}\}\Big),\quad J_2=\frac{1}{4}\Big(\{A^{R}_{-},A^{R}_{+}\}-\{A^{L}_{-},A^{L}_{+}\}\Big),\\
\label{Circular-Symmetries-2}
J_3=\frac{1}{4}\Big(\{A^{L}_{-},A^{R}_{+}\}+\{A^{L}_{+},A^{R}_{-}\}\Big),\quad \mathcal{H}=\frac{1}{2}\Big(\{A^{L}_{-},A^{L}_{+}\}+\{A^{R}_{-},A^{R}_{+}\}\Big).
\end{gather}
\end{subequations}
Using the above formulas and the actions of the circular operators $A^{L/R}_{\pm}$, the matrix elements of the $sd(2)$ generators in the circular basis can be computed; they are given below for $J_2$ and $J_3$. The action of the Hamiltonian $\mathcal{H}$ is
\begin{align*}
\mathcal{H}\ket{n_L,n_R}=(n_L+n_R+\mu_x+\mu_y+1)\ket{n_L,n_R}.
\end{align*}
It is clear that the generators preserve the subspace spanned by the basis vectors $\{\ket{n_L,n_R}\,\rvert\,n_L+n_R=N\}$. As is seen from the action of $\mathcal{H}$, this corresponds to the space of degenerate eigenstates of $\mathcal{H}$ with energy $\mathcal{E}_{N}$. The properties of representations of the symmetry generators in the circular basis will now be used to derive the transition matrix from the circular basis to the Cartesian basis and to obtain the eigenvalues of $J_2$ algebraically.

\subsection{Transition matrix from the circular to the Cartesian basis }
We consider the $N+1$-dimensional energy eigenspace spanned by the circular basis vectors $\ket{n_L,n_R}$ with $n_L+n_R=N$ and redefine the basis vectors as follows
\begin{align*}
\mathcal{B}_{1}:=\{f_0=\ket{0_{L},N_{R}},\,f_1=\ket{1_{L},(N-1)_{R}},\ldots,f_{N}=\ket{N_{L},0_{R}}\}.
\end{align*}
On this basis, a direct computation shows that the generator $J_3$ has the action
\begin{align}
\label{Dompe}
J_3 f_{n}=\frac{1}{2}\Big\{(N-n)f_{n+1}+\xi f_{n}+n f_{n-1}\Big\},
\end{align}
where we have defined
\begin{equation*}
\xi=\mu_x-\mu_y.
\end{equation*}
Since $J_3$ is diagonal in the Cartesian basis and tridiagonal in the circular basis, the two bases are related by a similarity transformation involving orthogonal polynomials.

Let us consider the decomposition of the Cartesian basis vector $v_{j}^{(\mu_x,\mu_y)}$ of $V_{N}^{(\mu_x,\mu_y)}$ on the circular basis
\begin{align}
\label{Ultra}
v_{j}^{(\mu_x,\mu_y)}=\sum_{n=0}^{N}C_{n}(j)f_{n},
\end{align}
where $j\in \{0,\ldots,N\}$. Acting on both sides of \eqref{Ultra} with $J_3$ and using \eqref{J3-Cartesian} and \eqref{Dompe}, one arrives at the following recurrence relation satisfied by the expansion coefficients $C_n(j)$:
\begin{align*}
(2j-N)C_{n}(j)=(n+1)C_{n+1}(j)+(N-n+1)C_{n-1}(j),
\end{align*}
with $C_{-1}=0$. Upon factoring out the initial value 
\begin{align*}
C_{n}(j)=C_{0}(j)P_{n}(j),
\end{align*}
we obtain the recurrence relation
\begin{align}
\label{3-Term}
(2j-N)P_{n}(j)=(n+1)\,P_{n+1}(j)+(N-n+1)\,P_{n-1}(j),
\end{align}
where $P_0(j)=1$. It follows from \eqref{3-Term} that $P_{n}(x)$ is a polynomial of degree $n$ in $x$. Upon substituting $P_{n}(j)=\widehat{P}_{n}(j)/n!$, we obtain the normalized recurrence relation
\begin{align*}
(j-N/2)\widehat{P}_{n}(j)=\widehat{P}_{n+1}(j)+\frac{1}{4}n(N-n+1)\widehat{P}_{n-1}(j).
\end{align*}
It is directly seen that the polynomials $\widehat{P}_{n}(j)$ are the monic Krawtchouk polynomials $K_{n}(x;p,N)$ \cite{Koekoek-2010} with parameter $p=1/2$ and variable $x$ evaluated at $x=j$. We thus have
\begin{align*}
C_{n}(j)=C_0(j)K_{n}(j;1/2,N),
\end{align*}
where the constant $C_0(j)$ can be chosen to ensure the unitarity of the transition matrix by using the orthogonality relation of the Krawtchouk polynomials. Despite the differences that the Dunkl and standard harmonic oscillators exhibit, the relations between the circular and Cartesian bases are identical in both cases.
\subsection{Matrix representation of $J_2$ and spectrum}
The circular representation space can be used to derive the spectrum of the symmetry operator $J_2$. To exhibit the structure of $J_2$, we introduce the following notation for the basis vectors:
\begin{align*}
\ket{n_L,n_R}=\ket{\ell,\pm},
\end{align*}
where 
\begin{align*}
\ell=\lfloor |n_L-n_R|/2\rfloor,\qquad \pm =\mathrm{sign}\,(n_R-n_L),
\end{align*}
where $\lfloor x\rfloor$ is the floor function. We adopt the convention that 
\begin{align*}
\mathrm{sign}\,0=- 1
\end{align*}
for convenience. We denote by $\mathcal{B}_2$ the circular basis such that $n_L+n_R=N$ with the ordering
\begin{align*}
\mathcal{B}_{2}=\{\ket{0,+},\ket{0,-},\ket{1,+},\ket{1,-},\ldots,\}
\end{align*}
As an example, consider the case $N=3$. The basis reads
\begin{align*}
\mathcal{B}_2= \{\ket{0,+},\ket{0,-},\ket{1,+},\ket{1,-}\}
\end{align*}
and corresponds to the following ordering of the standard circular basis vectors $\ket{n_L,n_R}$:
\begin{align*}
\mathcal{B}_2=\{\ket{1,2},\ket{2,1},\ket{0,3},\ket{3,0}\}.
\end{align*}
For $N=4$, one has
\begin{align*}
\mathcal{B}_{2}=\{\ket{0,-},\ket{1,+},\ket{1,-},\ket{2,+},\ket{2,-}\},
\end{align*}
which corresponds to
\begin{align*}
\mathcal{B}_2=\{\ket{2,2},\ket{1,3},\ket{3,1},\ket{0,4},\ket{4,0}\}.
\end{align*}
Using the action of the operators $A_{\pm}^{L/R}$ and the formulas \eqref{Circular-Symmetries-1}, \eqref{Circular-Symmetries-2}, the matrix representation of $J_2$ in the circular basis $\mathcal{B}_2$ is derived in a straightforward manner. We find that for $N$ even, the $N+1$-dimensional square matrix representing $J_2$ in the basis $\mathcal{B}_2$ is block upper-triangular with all blocks $2\times 2$ in addition to a row of $1\times 2$ blocks. The matrix reads
\small
\begin{align}
\label{MAT-1}
[J_2]_{\mathcal{B}_2}=
\begin{pmatrix}
0& \omega_1&\omega_2&\cdots &&\omega_m\\
& \Gamma_1 & \Omega_1 &\Omega_2&\cdots& \Omega_{m-1}\\
& & \Gamma_2 & \Omega_1 &\cdots &\Omega_{m-2}\\
& & & \ddots &&\vdots\\
& & & & \Gamma_{m-1} & \Omega_{1}\\
&&&&& \Gamma_{m}
\end{pmatrix},
\end{align}
\normalsize
with $m=N/2$ and where we have
\small
\begin{align}
\label{Detail}
\Gamma_{k}=
\begin{pmatrix}
k+\zeta/2 & -\zeta/2\\
\zeta/2 & -k-\zeta/2
\end{pmatrix},
\quad 
\Omega_{k}=
\begin{cases}
\begin{pmatrix}
-\xi & \xi\\
-\xi & \xi
\end{pmatrix} & \text{$k$ odd,}\\
\begin{pmatrix}
\zeta & -\zeta \\
\zeta & -\zeta
\end{pmatrix}& \text{$k$ even,}
\end{cases},
\end{align}
\normalsize
with $\omega_{k}$ corresponding to the lower part of $\Omega_k$. We have taken
\begin{align*}
\zeta=\mu_x+\mu_y,\qquad \xi=\mu_x-\mu_y.
\end{align*}
In the $N$ odd case, one obtains
\small
\begin{align}
\label{MAT-2}
[J_2]_{\mathcal{B}_2}=
\begin{pmatrix}
 \widetilde{\Gamma}_0 & \widetilde{\Omega}_1 &\widetilde{\Omega}_2&\cdots& \widetilde{\Omega}_{m}\\
 & \widetilde{\Gamma}_1 & \widetilde{\Omega}_1 &\cdots &\widetilde{\Omega}_{m-1}\\
 & & \ddots &&\\
 & & & \widetilde{\Gamma}_{m-1} & \widetilde{\Omega}_{1}\\
&&&& \widetilde{\Gamma}_{m}
\end{pmatrix},
\end{align}
\normalsize
with $m=(N-1)/2$ and where
\small
\begin{align}
\label{Detail-2}
\widetilde{\Gamma}_{k}=
\begin{pmatrix}
(2k+1+\zeta)/2 & \xi/2\\
-\xi/2 & -(2k+1+\zeta)/2
\end{pmatrix},
\quad 
\widetilde{\Omega}_{k}=
\begin{cases}
\begin{pmatrix}
-\xi & -\zeta\\
\zeta & \xi
\end{pmatrix} & \text{$k$ odd,}\\
\begin{pmatrix}
\zeta & \xi \\
-\xi & -\zeta
\end{pmatrix}& \text{$k$ even,}
\end{cases}.
\end{align}
\normalsize
Since in both cases the matrices representing $J_2$ are block upper-triangular, it follows from elementary linear algebra that the set of eigenvalues of $J_2$ is the union of the sets of eigenvalues of each diagonal block $\Gamma_{k}$ or $\widetilde{\Gamma}_k$. By the direct diagonalization of the $2\times 2$ diagonal blocks, we obtain that when $N$ is even, the eigenvalues of $J_2$ are given by
\begin{align*}
\lambda^{\pm}_{k}=\pm \sqrt{k(k+\mu_x+\mu_y)}, \qquad k=0,\ldots,m,
\end{align*}
where $m=N/2$ and where the eigenvalue $\lambda_0^{-}=0$ is non-degenerate. When $N$ is odd, the spectrum of $J_2$ has the form
\begin{align*}
\lambda^{\pm}_{k}=\pm\sqrt{(k+\mu_x+1/2)(k+\mu_y+1/2)},\qquad k=0,\ldots,m',
\end{align*}
where $m'=(N-1)/2$. These eigenvalues are indeed the eigenvalues of $J_2$ that were obtained in the first part \cite{Genest-2012-4} by solving the differential equation arising from the realization of $J_2$ in terms of Dunkl derivatives; here they have been obtained in a purely algebraic manner. It is seen from the matrices \eqref{MAT-1}, \eqref{MAT-2} that when $\mu_x=\mu_y=0$, the matrices representing $J_2$ are diagonal. This corresponds to the standard result for the harmonic oscillator, where the circular basis is the eigenbasis of the symmetry $J_2$.

Given that in the case of the Schwinger-Dunkl algebra $sd(2)$, the circular basis does not diagonalize $J_2$ directly, it is of interest to inquire about the eigenvectors of $J_2$ in this basis. This is the subject of the next two sections.
\section{Diagonalization of $J_2$: the $N$ even case}
This section is devoted to the computation of the eigenvectors of $J_2$ in the circular basis for the $N$ even case. To perform the calculation, we shall make use of an auxiliary operator $\mathcal{Q}$ whose eigenvectors have been related to those of $J_2$ in the previous paper \cite{Genest-2012-4}. The evaluation of the eigenvectors of $\mathcal{Q}$ is somewhat involved and consequently it is instructive to first expose the main steps of the computation. 

Firstly, the structure of $\mathcal{Q}$ will be used to reduce the eigenvalue problem to a system of recurrence relations for the components of the eigenvectors. Secondly, using generating functions, the recurrence system will be transformed into a system of differential equations and the solutions will be expressed in terms of Heun polynomials. Thirdly, using well-known properties of Heun functions, the explicit expressions for the components of the eigenvectors will be obtained in terms of a special case of complementary Bannai-Ito polynomials \cite{Genest-2012-1} which correspond to para-Krawtchouk polynomials \cite{Vinet-2012-3}. Lastly, the relation between the eigenvectors of $J_2$ and $\mathcal{Q}$ obtained in the first paper \cite{Genest-2012-4} will be used to write the final expression for the eigenvectors of $J_2$ in the circular basis.

\subsection{The operator $\mathcal{Q}$ and its simultaneous eigenvalue equation}
The operator $\mathcal{Q}$ has been used in the first part \cite{Genest-2012-4} to obtain the overlap coefficients between the wavefunctions in Cartesian and polar coordinates of the 2D Dunkl oscillator \cite{Genest-2012-2}. It is defined in terms of $J_2$ through the relation
\begin{align}
\label{Def-Q}
\mathcal{Q}=-2iJ_2R_{x}-\mu_xR_{y}-\mu_y R_x-(1/2)R_xR_y.
\end{align}
Given the action \eqref{Action-Reflection} of the reflections operators, it is seen that $R_x$, $R_y$ have the following matrix representation in the circular basis $\mathcal{B}_2$:
\begin{align}
\label{Reflection-Matrix}
R_x=R_y=\mathrm{diag}(1,\sigma_1,\sigma_1,\cdots,\sigma_1),\quad \sigma_1=\begin{pmatrix}0 & 1\\ 1 &0\end{pmatrix}.
\end{align}
Using the formula \eqref{Reflection-Matrix} for the reflections and the formulas \eqref{MAT-1}, \eqref{Detail} for the expression of $J_2$ in the $N$ even case, one obtains from \eqref{Def-Q}
\small
\begin{align*}
[\mathcal{Q}]_{\mathcal{B}_2}=
\begin{pmatrix}
\phi_0 & \delta_1 & \delta_2 & \cdots & \delta_{m}\\
& \Phi_1 & \Delta_1 & \cdots&  \Delta_{m-1}\\
& & \ddots & &\vdots\\
&&&\Phi_{m-1}&\Delta_{1}\\
&&&& \Phi_{m},
\end{pmatrix},
\end{align*}
\normalsize
with $m=N/2$, $\phi_0=-\zeta-1/2$ and where
\small
\begin{align}
\label{Homier}
\Phi_{k}=
\begin{pmatrix}
i\zeta-1/2 & -2ik-(1+i)\zeta\\
2ik-(1-i)\zeta & -i\zeta-1/2
\end{pmatrix},
\quad
\Delta_{m}=
\begin{cases}
\begin{pmatrix}
-2i\xi & 2i\xi\\
-2i\xi & 2i\xi
\end{pmatrix} & m\,\text{odd}\\
\begin{pmatrix}
2i\zeta & -2i\zeta\\
2i\zeta& -2i\zeta
\end{pmatrix} & m\,\text{even}
\end{cases}.
\end{align}
\normalsize
The $1\times 2$ blocks $\delta_{i}$ correspond to the lower part of the blocks $\Delta_{i}$. From the block upper-triangular structure, it follows that the eigenvalues $\nu_{k}^{\pm}$ of $\mathcal{Q}$ are given by
\begin{align}
\nu_{k}^{+}=2k+\zeta-1/2,\quad \nu_{k}^{-}=-(2k+\zeta+1/2),\quad k=1,\ldots,m,
\end{align}
and we also have $\nu_{0}^{-}=-\zeta-1/2$. Let us denote by $\ket{k,\pm}_{\mathcal{Q}}$ the eigenvectors of $\mathcal{Q}$ with eigenvalues $\nu_k^{\pm}$. We wish to evaluate the components of these eigenvectors in the circular basis. We define their expansion in the circular basis by
\begin{gather}
\label{Components}
\ket{k,+}_{\mathcal{Q}}=\sum_{\substack{\ell=0\\\sigma=\pm}}^{m}u_{\ell}^{\sigma}(k)\ket{\ell,\sigma},\qquad \ket{k,-}_{\mathcal{Q}}=\sum_{\substack{\ell=0\\\sigma=\pm}}^{m}v_{\ell}^{\sigma}(k)\ket{\ell,\sigma},
\end{gather}
for $k=1,\ldots,m$ and where the vectors $\ket{\ell,\pm}$ are vectors of the circular basis $\mathcal{B}_2$. It is clear from the matrix representation of $\mathcal{Q}$ that $\ket{0,-}_{\mathcal{Q}}=\ket{0,-}$ and thus $v_{0}^{-}(0)=1$.

We shall study the simultaneous eigenvalue equation for the operator $\mathcal{Q}$. Since the matrix representing $\mathcal{Q}$ is block upper-triangular, the matrix of eigenvectors will have the same structure. We define the matrix of eigenvectors of $\mathcal{Q}$ as follows:
\small
\begin{align*}
W=
\begin{pmatrix}
1 & \widetilde{V}_{01} & \widetilde{V}_{02} &\scriptscriptstyle\cdots & \widetilde{V}_{0m}\\
& V_{11} & V_{12} & \scriptscriptstyle\cdots  & V_{1m}\\
& & V_{22} & & \vdots \\
& & & \scriptscriptstyle \ddots& V_{m-1m}\\
&&&& V_{mm}
\end{pmatrix},
\end{align*}
\normalsize
where
\small
\begin{align*}
V_{\ell,k}=
\begin{pmatrix}
u_{\ell}^{+}(k) & v_{\ell}^{+}(k)\\
u_{\ell}^{-}(k) & v_{\ell}^{-}(k)
\end{pmatrix},
\end{align*}
\normalsize
and where $\widetilde{V}_{\ell k}$ is the $1\times 2$ block corresponding to the lower part of $V_{\ell k}$. The simultaneous eigenvalue equation for the matrix $[\mathcal{Q}]_{\mathcal{B}_2}$ can be written as
\begin{align}
\label{Eigen}
W\cdot L=[\mathcal{Q}]_{\mathcal{B}_2}\cdot W,
\end{align}
with
\small
\begin{align*}
L=\mathrm{diag}(\nu_0^{(-)},\,\Lambda_1,\cdots,\Lambda_m),\quad \Lambda_{k}=
\begin{pmatrix}
2k+\zeta-1/2 & 0\\
0 & -2k-\zeta-1/2
\end{pmatrix}.
\end{align*}
\normalsize
As will be seen, the components \eqref{Components} of the eigenvectors of $\mathcal{Q}$ can be derived from the eigenvalue equation \eqref{Eigen} by solving the associated system of recurrence relations.
\subsection{Recurrence relations}
It will prove convenient to consider the two sectors corresponding to the eigenvalues $\nu_{k}^{+}$, $\nu_{k}^{-}$ separately. In block form, for $\ell=1,\ldots,m$ and $k=1,\ldots,m$, the eigenvalue equation \eqref{Eigen} can be written in the form
\begin{align}
\label{Eigen-2}
V_{\ell k}\Lambda_{k}=\Phi_{\ell}V_{\ell k}+\sum_{j=1}^{k-\ell}\Delta_{j}V_{j k},
\end{align}
with $\Phi_{\ell}$ and $\Delta_{j}$ given in \eqref{Homier} and where the range of the sum has been determined by the structure of the eigenvector matrix $W$.
\subsubsection{The $\nu^{+}_k$ eigenvalue sector}\hfill\bigskip

\noindent
We consider the eigenvectors $\ket{k,+}_{\mathcal{Q}}$ of $\mathcal{Q}$ with the expansion
\begin{align}
\label{Expansion-Even-1}
\ket{k,+}_{\mathcal{Q}}=\sum_{\substack{\ell=0\\\sigma=\pm}}^{k}u_{\ell}^{\sigma}(k)\ket{\ell,\sigma},
\end{align}
and associated to the eigenvalue $\nu^{+}_{k}=2k+\zeta-1/2$. It is understood that $u_{0}^{+}(k)$ does not belong to this decomposition. For $\ell=1,\ldots,m$, it directly seen that the eigenvalue equation \eqref{Eigen-2} is equivalent to the following system of recurrence relations:
\small
\begin{subequations}
\label{Recu-Syst}
\begin{align}
[2k+(1-i)\zeta]u_{\ell}^{+}&=[-2i\ell-(1+i)\zeta]u_{\ell}^{-}-2i\sum_{j=\ell+1}^{k}\{(-1)^{j-\ell}\mu_x+\mu_y\}B_{j},\\
[2k+(1+i)\zeta]u_{\ell}^{-}&=[2i\ell-(1-i)\zeta]u_{\ell}^{+}-2i\sum_{j=\ell+1}^{k}\{(-1)^{j-\ell}\mu_x+\mu_y\}B_{j},
\end{align}
\end{subequations}
\normalsize
where we have defined 
\begin{equation*}
B_{j}=u^{-}_{j}-u^{+}_{j}
\end{equation*}
and where the explicit dependence of the components $u_{\ell}^{\pm}$ on $k$ has been dropped for notational convenience. The case $\ell=0$ is treated below. The system of recurrence relations \eqref{Recu-Syst} is ''reversed'': the values of $u_{i}^{\pm}$ are obtained from the values of  $u_{j}^{\pm}$ with $i<j$ and $j<p$. The terminating conditions are at $\ell=k$. In this case \eqref{Recu-Syst} reduces to
\begin{subequations}
\label{Recu-Initial}
\begin{align}
[2k+(1-i)\zeta]u_{k}^{+}&=[-(2k)i-(1+i)\zeta]u_{k}^{-},\\
[2k+(1+i)\zeta]u_{k}^{-}&=[(2k)i-(1-i)\zeta]u_{k}^{+}.
\end{align}
\end{subequations}
In accordance to the system \eqref{Recu-Initial}, we choose the following terminating conditions
\begin{align*}
u_{k}^{+}=-i,\quad u_{k}^{-}=1.
\end{align*}
Upon introducing 
\begin{align*}
A_{j}=u_{j}^{+}+u_{j}^{-},
\end{align*}
the system \eqref{Recu-Syst} is directly seen to be equivalent to
\begin{subequations}
\begin{align}
[k+\zeta]A_{\ell}&=-i(\ell+\zeta)B_{\ell}-2i\sum_{j=\ell+1}^{k}\{(-1)^{j-\ell}\mu_x+\mu_y\}B_{j},\\
[k]B_{\ell}&=i\ell A_{\ell}.
\end{align}
\end{subequations}
The above system accounts for the $\ell=0$ case. Indeed, it is seen that $B_{0}=0$ and hence $u_{0}^{-}=A_{0}/2$.
These equations can be simplified by factoring out the terminating conditions 
\begin{equation*}
A_{\ell}=\alpha_0\widehat{A}_{\ell},\qquad  B_{\ell}=\beta_0\widehat{B}_{\ell},
\end{equation*}
where $\alpha_0=(1-i)$ and $\beta_0=(1+i)$. It is seen that the normalized components $\widehat{A}_{\ell}$, $\widehat{B}_{\ell}$ are real and satisfy the system
\begin{subequations}
\label{Recu-Syst-3}
\begin{align}
(k+\zeta)\widehat{A}_{\ell}&=[\ell+\zeta]\widehat{B}_{\ell}+2\sum_{j=\ell+1}^{k}\{(-1)^{j-\ell}\mu_x+\mu_y\}\widehat{B}_{j},\\
k\widehat{B}_{\ell}&=\ell \widehat{A}_{\ell},
\end{align}
\end{subequations}
with the terminating conditions $\widehat{A}_{k}=\widehat{B}_{k}=1$. The system \eqref{Recu-Syst-3} can be simplified by introducing the reversed components $\widetilde{a}_{\ell}=\widehat{A}_{k-\ell}$ and $\widetilde{b}_{\ell}=\widehat{B}_{k-\ell}$. Using the index $n$, the system takes the usual form
\begin{subequations}
\label{Recu-Syst-4}
\begin{align}
(k+\zeta)\widetilde{a}_{n}&=(k-n+\zeta)\widetilde{b}_{n}+2\sum_{\alpha=0}^{n-1}\{(-1)^{n+\alpha}\mu_x+\mu_y\}\widetilde{b}_{\alpha}\\
k\,\widetilde{b}_{n}&=(k-n)\widetilde{a}_{n},
\end{align}
with the initial conditions $\widetilde{a}_0=\widetilde{b}_0=1$. Hence the components $u_{\ell}^{\pm}(k)$ of the eigenvector $\ket{k,\pm}_{\mathcal{Q}}$ of the operator $\mathcal{Q}$ have the expression
\end{subequations}
\small
\begin{equation}
\label{Plus-Sector-Even}
u_{\ell}^{-}(k)=\frac{\alpha_0\widetilde{a}_{k-\ell}+\beta_0\widetilde{b}_{k-\ell}}{2},\quad u_{\ell}^{+}(k)=\frac{\alpha_0\widetilde{a}_{k-\ell}-\beta_0\widetilde{b}_{k-\ell}}{2},
\end{equation}
\normalsize
where $\widetilde{a}_{n}$ and $\widetilde{b}_{n}$ are the unique solutions to the system \eqref{Recu-Syst-4}.
\subsubsection{The $\nu^{-}_k$ eigenvalue sector}\hfill\bigskip

\noindent
We consider the eigenvectors $\ket{k,-}_{\mathcal{Q}}$ corresponding to the eigenvalue $\nu^{-}_{k}$ of $\mathcal{Q}$ with the circular basis expansion
\begin{align*}
\ket{k,-}_{\mathcal{Q}}=\sum_{\substack{\ell=0\\ \sigma=\pm}}^{k}v_{\ell}^{\sigma}(k)\ket{\ell,\sigma},
\end{align*}
and associated eigenvalue $\nu_{k}^{-}=-2k-\zeta-1/2$. An analysis similar to the preceding one shows that the components $v_{\ell}^{\pm}(k)$ differ from the components $u_{\ell}^{\pm}(k)$ only by their terminating conditions. Again choosing $v_{k}^{-}(k)=1$, we find
\small
\begin{align*}
v_{k}^{+}(k)=\frac{(1+i)k+\zeta}{(1-i)k-\zeta},\qquad v_{k}^{-}(k)=1.
\end{align*}
\normalsize
This yields
\small
\begin{align}
\label{Minus-Sector-Even}
v_{\ell}^{-}(k)=\frac{\gamma_0\widetilde{a}_{k-\ell}+\epsilon_0 \widetilde{b}_{k-\ell}}{2},\quad v_{\ell}^{+}(k)=\frac{\gamma_0\widetilde{a}_{k-\ell}+\epsilon_0 \widetilde{b}_{k-\ell}}{2},
\end{align}
\normalsize
where
\begin{align*}
\gamma_0=\frac{2i(k+\zeta)}{(1+i)k+i\zeta},\quad \epsilon_0=\frac{2k}{(1+i)k+i\zeta},
\end{align*}
and where $v_{0}^{-}=\gamma_0\widetilde{a}_{k}$.
\normalsize

\noindent

\subsection{Generating function and Heun polynomials}
We have seen that the evaluation of the components of the eigenvectors of $\mathcal{Q}$ in the circular basis depends on the solution of the recurrence system \eqref{Recu-Syst-4}. As it turns out, an explicit solution for $\widetilde{a}_n(k)$ and $\widetilde{b}_{n}(k)$ can be obtained using generating functions. 

We introduce the ordinary generating functions
\begin{align*}
\widetilde{A}(z)=\sum_{n\geqslant 0}\widetilde{a}_{n}z^{n},\qquad \widetilde{B}(z)=\sum_{n\geqslant 0}\widetilde{b}_{n}z^{n}.
\end{align*}
We shall make use of the elementary identities
\small
\begin{subequations}
\label{GenFun-Iden}
\begin{gather}
z\pd_{z}\widetilde{A}(z)=\sum_{n\geqslant0}n\widetilde{a}_{n}z^{n},\qquad
(1-z)^{-1}\widetilde{A}(z)=\sum_{n\geqslant0}\left(\sum_{0\leqslant k\leqslant n} \widetilde{a}_{k}\right)z^{n},\\
(1+z)^{-1}\widetilde{A}(z)=\sum_{n\geqslant0}\left(\sum_{0\leqslant k\leqslant n} (-1)^{k+n}\widetilde{a}_{k}\right)z^{n}.
\end{gather}
\end{subequations}
\normalsize
Using the above identities, it is easily seen that the system of recurrence relations \eqref{Recu-Syst-4} for the quantities $\widetilde{a}_n$, $\widetilde{b}_n$ is equivalent to the following system of differential equations for the generating functions $\widetilde{A}(z)$, $\widetilde{B}(z)$:
\small
\begin{subequations}
\label{Diff-Syst-1}
\begin{align}
(k+\zeta)\widetilde{A}(z)&=(k-\zeta-z\pd_{z})\widetilde{B}(z)+\frac{2\mu_x}{1+z}\widetilde{B}(z)+\frac{2\mu_y}{1-z}\widetilde{B}(z), \\
k\widetilde{B}(z)&=(k-z\pd_{z})\widetilde{A}(z).
\end{align}
\end{subequations}
\normalsize
By direct substitution, we find that the generating function $\widetilde{A}(z)$ satisfies the second-order differential equation
\small
\begin{align}
\label{Heun-1}
\widetilde{A}''(z)+\left(\frac{1-2k-\zeta}{z}+\frac{2\mu_y}{z-1}+\frac{2\mu_x}{1+z}\right)\widetilde{A}'(z)+\left(\frac{-2k\zeta z+2k\xi}{z(z-1)(z+1)}\right)\widetilde{A}(z)=0.
\end{align}
\normalsize
This corresponds to Heun's differential equation \cite{NIST:DLMF,Ronveaux-1995}. The general form of the Heun differential equation is
\small
\begin{align}
\label{Heun}
w''(z)+\left(\frac{\gamma}{z}+\frac{\delta}{z-1}+\frac{\epsilon}{z-a}\right)w'(z)+\frac{\alpha\beta z-q}{z(z-1)(z-a)}w(z)=0,
\end{align}
\normalsize
with $\alpha+\beta+1=\gamma+\delta+\epsilon$. Comparing \eqref{Heun} with \eqref{Heun-1}, we thus write
\begin{align}
\label{Generating-A}
\widetilde{A}(z)=H\ell(a,q;\alpha,\beta,\gamma,\delta,z)
\end{align}
with the parameters
\begin{subequations}
\label{Parameters}
\begin{gather}
a=-1,\quad q=2k(\mu_y-\mu_x),\quad \alpha=-2k,\\
\beta=\mu_x+\mu_y,\quad \gamma=1-2k-\mu_x-\mu_y,\quad \delta=2\mu_y.
\end{gather}
\end{subequations}
The function $H\ell(a,q;\alpha,\beta,\gamma,\delta)$ denotes the solution to \eqref{Heun} that corresponds to the exponent $0$ at $z=0$ and assumes the value $1$ at that point. This is obviously the case of $\widetilde{A}(z)$. It will be seen that $\widetilde{A}(z)$ is in fact a polynomial of degree $2k$, and hence that the Heun function \eqref{Generating-A} is in fact a \emph{Heun polynomial}. Given the system \eqref{Diff-Syst-1}, we also have
\begin{align*}
\widetilde{B}(z)=k^{-1}(k-z\pd_{z})\widetilde{A}(z).
\end{align*}
\subsection{Expansion of Heun polynomials in the complementary Bannai-Ito polynomials}
The well-studied properties of Heun functions can be used to obtain a closed form formula for the coefficients $\widetilde{a}_{n}$ and hence for $\widetilde{b}_{n}$. In what follows, it will be shown that the Heun polynomial in \eqref{Generating-A} can be expanded in terms of a special case of the complementary Bannai-Ito polynomials corresponding to the para-Krawtchouk polynomials.

Consider the solution $H\ell(a,q;\alpha,\beta,\gamma,\delta)$ to the equation \eqref{Heun} and its Maclaurin expansion
\begin{align*}
H\ell(a,q;\alpha,\beta,\gamma,\delta)=\sum_{n=0}^{\infty}c_{n}z^{n},
\end{align*}
where $c_{-1}=0$, $c_0=1$. The coefficients $c_n$ obey the three-term recurrence relation \cite{NIST:DLMF,Ronveaux-1995}
\begin{align*}
R_{n}c_{n+1}-(Q_{n}+q)c_{n}+P_{n}c_{n-1}=0,
\end{align*}
where 
\begin{subequations}
\label{Recurrence-Coeff}
\begin{gather}
R_{n}=a(n+1)(n+\gamma),\quad Q_{n}=n\big[(n-1+\gamma)(1+a)+a\delta+\epsilon\big],\\
P_{n}=(n-1+\alpha)(n-1+\beta).
\end{gather}
\end{subequations}
The identification of $\widetilde{A}(z)$ as a Heun function enables one to reduce the evaluation of $\widetilde{a}_{n}$ to the solution of a three-term recurrence relation. It is seen that with the choice of paramaters \eqref{Parameters}, the expansion coefficients $c_{n}$ of the Heun function $\widetilde{A}(z)$ truncate at degree $k=2n$. Hence $\widetilde{A}(z)$ is a polynomial of degree $2k$ in $z$. For convenience, we use the symbol $\mathcal{P}_{n}=\widetilde{a}_{n}$ in the computations to follow. Using the parameters \eqref{Parameters} in the recurrence coefficients \eqref{Recurrence-Coeff} for the expansion coefficients in
\begin{align*}
\widetilde{A}(z)=\sum_{n\geqslant 0}\mathcal{P}_{n}z^{n},
\end{align*}
we can obtain the recurrence relation for $\mathcal{P}_{n}$. Upon setting $N=2k$ and dividing by $(N-2n)$ we find that $\mathcal{P}_{n}(\xi)$ is a symmetric polynomial of degree $n$ in the variable $\xi$ obeying the recurrence relation
\begin{align*}
\sigma_{n+1}\mathcal{P}_{n+1}(\xi)-\kappa_{n}\mathcal{P}_{n-1}(\xi)=\xi\,\mathcal{P}_{n}(\xi),
\end{align*}
with $\mathcal{P}_{-1}=0$, $\mathcal{P}_{1}=1$ and where
\begin{gather*}
\sigma_{n}=\frac{n(N+\zeta-n)}{(2n-N-2)},\quad \kappa_{n}=\frac{(N+1-n)(n+\zeta-1)}{(2n-N)}.
\end{gather*}
As $k$ takes integer values, it is seen that when $k=n$, a singularity appears in the recurrence relation. Notwithstanding this, we proceed with the computation; the effect of the pole in $k=n$  on the results is treated below.

Introducing the monic polynomials $\mathcal{P}_{n}(\xi)=\frac{\widehat{\mathcal{P}}_{n}(\xi)}{\sigma_{1}\cdots \sigma_{n}}$, the recurrence relation becomes
\begin{align}
\label{Recu-1}
\widehat{\mathcal{P}}_{n+1}(\xi)+u_{n}\widehat{\mathcal{P}}_{n-1}(\xi)=\xi\,\widehat{\mathcal{P}}_{n}(\xi),
\end{align}
where 
\begin{align*}
u_{n}=-\frac{n(N+1-n)(N-n+\zeta)(n+\zeta-1)}{(N-2n)(N-2n+2)}.
\end{align*}
The monic polynomials $\widehat{\mathcal{P}}_{n}(\xi)$ can be identified with the complementary Bannai-Ito polynomials (CBI).

The monic CBI polynomials \cite{Genest-2012-1,Vinet-2012-2}, denoted $I_{n}(x;\rho_1,\rho_2,r_1,r_2)$, obey the recurrence relation
\begin{align}
\label{CBI-REcu}
I_{n+1}(x)+(-1)^{n}\rho_2 I_{n}(x)+\tau_{n}I_{n-1}(x)=x I_{n}(x),
\end{align}
where 
\small
\begin{subequations}
\label{CBI-Coeff}
\begin{gather}
\tau_{2n}=-\frac{n(n+\rho_1-r_1+1/2)(n+\rho_1-r_2+1/2)(n-r_1-r_2)}{(2n+g)(2n+g+1)},
\\
\tau_{2n+1}=-\frac{(n+g+1)(n+\rho_1+\rho_2+1)(n+\rho_2-r_1+1/2)(n+\rho_2-r_2+1/2)}{(2n+g+1)(2n+g+2)}
\end{gather}
\end{subequations}
\normalsize
and with $g=\rho_1+\rho_2-r_1-r_2$. They have the hypergeometric representation
\begin{align*}
I_{2n}(x)=R_{n}(x),\quad I_{2n+1}(x)=(x-\rho_2)Q_{n}(x),
\end{align*}
where 
\begin{align}
\label{explicit}
R_{n}(x)&=\eta_{n}\,\pFq{4}{3}{-n,n+g+1,\rho_2+x,\rho_2-x}{\rho_1+\rho_2+1,\rho_2-r_1+1/2,\rho_2-r_2+1/2}{1},\\
Q_{n}(x)&=\iota_{n}\,\pFq{4}{3}{-n,n+g+2,\rho_2+x+1,\rho_2-x+1}{\rho_1+\rho_2+2,\rho_2-r_1+3/2,\rho_2-r_2+3/2}{1},
\end{align}
with
\small
\begin{align*}
\eta_{n}&=\frac{(\rho_1+\rho_2+1)_{n}(\rho_2-r_1+1/2)_{n}(\rho_2-r_2+1/2)}{(n+g+1)_{n}},\\
\iota_{n}&=\frac{(\rho_1+\rho_2+2)_{n}(\rho_2-r_1+3/2)_{n}(\rho_2-r_2+3/2)}{(n+g+2)_{n}},
\end{align*}
\normalsize
and where $(a)_{n}=(a)(a+1)\cdots(a+n-1)$ is the Pochhammer symbol.

Comparing the recurrence formulas \eqref{Recu-1} with \eqref{CBI-REcu} and \eqref{CBI-Coeff}, it is seen that the polynomials $\widehat{\mathcal{P}}_{n}(\xi)$ are monic CBI polynomials
\begin{align}
\widehat{P}_{n}(\xi)=I_{n}\left(\xi/2;\rho_1,\rho_2,r_1,r_2\right),
\end{align}
with
\begin{align}
\label{Para}
\rho_1=\frac{\zeta-2}{2},\quad \rho_2=0,\quad r_1=\frac{2k+\zeta}{2},\quad r_2=0.
\end{align}
The parametrization \eqref{Para} is a special case of CBI polynomials. This case corresponds to the para-Krawtchouk polynomials constructed in \cite{Vinet-2012-3} in the context of perfect state transfer in spin chains.

Since there is a singularity in the recurrence coefficients for the polynomials $\mathcal{P}_{n}(\xi)$, the correspondence between the polynomials $\mathcal{P}_{n}(\xi)$ and the CBI polynomials outlined above is valid only for $n=0,\ldots,k$ and hence the Heun polynomial $\widetilde{A}(z)$ generates only the first $k$ para-Krawtchouk polynomials. As is easily seen by induction, the recurrence relation \eqref{Recu-1} generates center-symmetric polynomials $\mathcal{P}_{n}$. Hence for $n>k$, we have $\mathcal{P}_{n}(\xi)=\mathcal{P}_{2k-n}(\xi)$. Putting the preceding results together, we write
\begin{align}
\label{Ultra-Dompe}
\widetilde{a}_{n}&=\frac{(-1)^{n}4^{n}}{n!}\frac{(k+1-n)_{n}}{(2k+\zeta-n)_{n}}I_{n}(\xi/2;\rho_1,\rho_2,r_1,r_2)
\end{align}
for $n\leqslant k$ and
\begin{align*}
\widetilde{a}_{n}&=\widetilde{a}_{2k-n},\quad n=k+1,\ldots,2k
\end{align*}
for $n=k+1\cdots, 2k$.

The hypergeometric expression of the CBI polynomials \eqref{explicit} provides an explicit formula for the coefficients $\widetilde{a}_{n}$ and the coefficients $\widetilde{b}_{n}$ are easily evaluated from the recurrence system \eqref{Recu-Syst-4}. Combining those results with the formulas \eqref{Plus-Sector-Even} and \eqref{Minus-Sector-Even} yields the expansion coefficients of the eigenvectors of  the operator $\mathcal{Q}$ in the circular basis. Note that these expansion coefficients only involve $\widetilde{a}_j$, $\widetilde{b}_j$ with $j=0,\ldots,k$ and hence only \eqref{Ultra-Dompe} is needed.

\subsection{Eigenvectors of $J_2$}
To obtain the expansion coefficients of the eigenvectors of $J_2$ in the circular basis, it is necessary to relate the eigenvectors of $J_2$ to those of $\mathcal{Q}$. This relation has been obtained in the previous paper \cite{Genest-2012-4}. In the present notation, we have
\begin{gather*}
\ket{k,+}_{\mathcal{Q}}=\frac{1}{\sqrt{2}}\left(\ket{k,+}_{J_2}-\omega_{k}\ket{k,-}_{J_2}\right),\\
\ket{k,-}_{\mathcal{Q}}=\frac{1}{\sqrt{2}}\left(\ket{k,+}_{J_2}+\omega_{k}\ket{k,-}_{J_2}\right)
\end{gather*}
where $\ket{k,\pm}_{J_2}$ are the eigenvectors of $J_2$ corresponding to the eigenvalues
\begin{align*}
\lambda_{\pm}=\sqrt{k(k+\zeta)},
\end{align*}
and where the coefficient $\omega_k$ is
\begin{align*}
\omega_{k}=\frac{\zeta-2i\sqrt{k(k+\zeta)}}{2k+\zeta}.
\end{align*}
The inverse relation, which allows to express the eigenvectors of $J_2$ in terms of the known eigenvectors of $\mathcal{Q}$ reads
\begin{align*}
\ket{k,+}_{J_2}&=\frac{1}{\sqrt{2}}\left(\ket{k,+}_{\mathcal{Q}}+\ket{k,-}_{\mathcal{Q}}\right),\\
\ket{k,-}_{J_2}&=\frac{-1}{\omega_{k}\sqrt{2}}\left(\ket{k,+}_{\mathcal{Q}}-\ket{k,-}_{\mathcal{Q}}\right).
\end{align*}

\subsection{The fully isotropic case}
We now consider the case $\mu_x=\mu_y=\mu$. This corresponds to a fully isotropic 2D Dunkl oscillator, where two "identical" parabosonic oscillators are combined. Returning to the system of differential equations \eqref{Diff-Syst-1} for the generating functions, one has
\begin{gather*}
(k+2\mu)\widetilde{A}(z)=(k-2\mu-z\pd_{z})\widetilde{B}(z)+\frac{4\mu}{1-z^2}\widetilde{B}(z),\\
k\widetilde{B}(z)=(k-z\pd_{z})\widetilde{A}(z).
\end{gather*}
Solving for $\widetilde{A}(z)$, we find
\begin{align*}
z(z^2-1)\widetilde{A}''(z)+\left(z^2(2\mu+1-2k)+2\mu+2k-1\right)\widetilde{A}'(z)-4k\mu z\widetilde{A}(z)=0.
\end{align*}
The solution corresponding to the initial value $\widetilde{a}_{0}=1$ is given by
\begin{align*}
\widetilde{A}(z)=\pFq{2}{1}{-k,\mu}{1-k-\mu}{z^2},
\end{align*}
and we also have
\begin{align*}
\widetilde{B}(z)=\widetilde{A}(z)-\,\frac{2k\mu z^2}{k+\mu-1}\,\pFq{2}{1}{1-k,1+\mu}{2-k-\mu}{z^2}.
\end{align*}
Hence in the isotropic case, the generating functions are simply the Jacobi polynomials. It follows from the hypergeometric generating function that
\begin{align*}
\widetilde{a}_{2n}=\frac{(-k)_{n}(\mu)_{n}}{(1-k-\mu)_{n}n!},\qquad \widetilde{a}_{2n+1}=0,
\end{align*}
and
\begin{align*}
\widetilde{b}_{2n}=\left(\frac{k-2n}{k}\right)\frac{(-k)_{n}(\mu)_{n}}{(1-k-\mu)_{n}n!},\qquad \widetilde{b}_{2n+1}=0.
\end{align*}
Thus it is seen that the in the fully isotropic case $\mu_x=\mu_y=\mu$, the formulas for the expansion coefficients of the eigenvectors of $\mathcal{Q}$ simplify substantially.
\section{Diagonalization of $J_2$: the $N$ odd case}
In this section, we obtain the expression for the eigenvectors of $J_2$ in the circular basis when $N=n_L+n_R$ is an odd integer. In spirit, the computation is similar to the $N$ even case presented in the previous section. We proceed along the same lines.
\subsection{The operator $\mathcal{Q}$ and its simultaneous eigenvalue equation}
The operator $\mathcal{Q}$ is defined by
\begin{align*}
\mathcal{Q}=-2iJ_2R_{x}-\mu_x R_{y}-\mu_{y}R_{x}-(1/2)R_{x}R_{y}.
\end{align*}
Given the action \eqref{Action-Reflection} of the reflections operators, it is seen that they have the following matrix representation in the circular basis $\mathcal{B}_2$ :
\begin{align*}
R_y=-R_{x}=\mathrm{diag}(\sigma_1,\ldots,\sigma_1),\quad 
\sigma_1=
\begin{pmatrix}
0 & 1\\
1 & 0
\end{pmatrix}.
\end{align*}
Using the matrix representation of $J_2$ in the circular basis $\mathcal{B}_2$ given in \eqref{MAT-2}, one finds
\small
\begin{align*}
[\mathcal{Q}]_{\mathcal{B}_2}=
\begin{pmatrix}
\widetilde{\Phi}_0 & \widetilde{\Delta}_1 & \widetilde{\Delta}_2 & \cdots &\widetilde{\Delta}_m\\
& \widetilde{\Phi}_1  & \widetilde{\Delta}_1 & \cdots&\widetilde{\Delta}_{m-1}\\
& & \ddots & &\\
&&&\widetilde{\Phi}_{m-1} & \widetilde{\Delta}_1\\
&&&&\widetilde{\Phi}_{m} 
\end{pmatrix},
\end{align*}
\normalsize
with $m=(N-1)/2$ and where
\small
\begin{align}
\label{Ultra-4}
\widetilde{\Phi}_{k}=
\begin{pmatrix}
1/2+i\xi & i(2k+\zeta+1)-\xi\\
-i(2k+\zeta+1)-\xi& 1/2-i\xi
\end{pmatrix},
\quad 
\widetilde{\Delta}_{m}=
\begin{cases}
\begin{pmatrix}
-2i\zeta & -2i\xi\\
2i\xi& 2i\zeta
\end{pmatrix}
 & \text{$m$ odd}\\
\begin{pmatrix}
2i\xi & 2i\zeta\\
-2i\zeta& -2i\xi
\end{pmatrix} & \text{$m$ even}
\end{cases}.
\end{align}
\normalsize
From the block upper-triangular structure, it follows that the eigenvalues $\nu_{k}^{\pm}$ of $\mathcal{Q}$ are
\begin{align*}
\nu_{k}^{+}=2k+\zeta+3/2,\qquad \nu_{k}^{-}=-(2k+\zeta+1/2),
\end{align*}
for $k=0,\ldots,m$. Although a different labeling has been used, it is directly checked that the eigenvalues of $\mathcal{Q}$ are the same for the $N$ even and $N$ odd case, except for the additional one. We denote the eigenvectors of $\mathcal{Q}$ corresponding to the eigenvalues $\nu_{k}^{\pm}$ by $\ket{k,\pm}_{\mathcal{Q}}$ and define their expansion in the circular basis by
\begin{align*}
\ket{k,+}_{\mathcal{Q}}=\sum_{\substack{\ell=0\\\sigma=\pm}}^{m}u_{\ell}^{\sigma}(k)\ket{\ell,\sigma},\qquad \ket{\ell,-}_{\mathcal{Q}}=\sum_{\substack{\ell=0\\\sigma=\pm}}^{m}v_{\ell}^{\sigma}(k)\ket{\ell,\sigma},
\end{align*}
for $k=0,\ldots,m$ and where the vectors $\ket{\ell,\pm}$ are the vectors of the circular basis $\mathcal{B}_2$. 

We shall once again study the simultaneous eigenvalue equation for the operator $\mathcal{Q}$. We define the matrix of eigenvectors
\small
\begin{align*}
W=
\begin{pmatrix}
V_{00} & V_{01} & \cdots & V_{0m}\\
& V_{11} & \cdots & V_{1m}\\
& & \ddots \\
& & & V_{mm}
\end{pmatrix},
\end{align*}
\normalsize
where 
\small
\begin{align*}
V_{\ell k}=
\begin{pmatrix}
v_{\ell}^{+}(k) & u_{\ell}^{+}(k)\\
v_{\ell}^{-}(k) & u_{\ell}^{-}(k)
\end{pmatrix}.
\end{align*}
\normalsize
The simultaneous eigenvalue equation for the matrix $[\mathcal{Q}]_{\mathcal{B}_2}$ reads
\begin{align}
\label{Ultra-2}
W\cdot L=[\mathcal{Q}]_{\mathcal{B}_2}W,
\end{align}
with
\small
\begin{align}
\label{Ultra-5}
L=\mathrm{diag}(\Lambda_0,\cdots, \Lambda_m),\qquad
\Lambda_{k}=
\begin{pmatrix}
\nu_k^{-}& 0 \\
0 & \nu_k^{+}
\end{pmatrix}.
\end{align}
\normalsize
As in section 4, the simultaneous equation \eqref{Ultra-2} will be shown to be equivalent to a system of recurrence relation for the components $u_{\ell}(k)^{\pm}$, $v_{\ell}^{\pm}(k)$ of the eigenvectors of $\mathcal{Q}$ in the circular basis.
\subsection{Recurrence relations}
We now construct the recurrence systems for the component of the eigenvectors of the operator $\mathcal{Q}$. For $k=0,\ldots,m$ and $\ell=0,\ldots,m$, the simultaneous equation \eqref{Ultra-2} takes the form
\begin{align}
\label{Ultra-3}
V_{\ell k}\Lambda_{k}=\widetilde{\Phi}_{\ell}V_{\ell k}+\sum_{j=1}^{k-\ell}\widetilde{\Delta}_{j}V_{j k}.
\end{align}

\subsubsection{The $\nu^{-}_k$ eigenvalue sector}\hfill\bigskip

\noindent
Let us begin by considering the eigenvectors $\ket{k,-}_{\mathcal{Q}}$ of $\mathcal{Q}$ corresponding to the eigenvalue $\nu_{k}^{-}$ and their expansion in the circular basis
\begin{align*}
\ket{k,-}_{\mathcal{Q}}=\sum_{\substack{\ell=0\\ \sigma=\pm}}^{k}v_{\ell}^{\sigma}(k)\ket{\ell,\sigma}.
\end{align*}
It is directly seen from \eqref{Ultra-3},\eqref{Ultra-4} and \eqref{Ultra-5} that we have
\small
\begin{subequations}
\label{Recu-Syst-5}
\begin{gather}
[2k+1+\zeta+i\xi]v_{\ell}^{+}=[\xi-i(2\ell+\zeta+1)]v_{\ell}^{-}-2i\mu_x \sum_{j=\ell+1}^{k}(-1)^{j-\ell}A_{j}-2i\mu_y\sum_{j=\ell+1}^{k}B_{j},\\
[2k+1+\zeta-i\xi]v_{\ell}^{-}=[\xi+i(2\ell+\zeta+1)]v_{\ell}^{+}+2i\mu_x\sum_{j=\ell+1}^{k}(-1)^{j-\ell}A_{j}-2i\mu_y\sum_{j=\ell+1}^{k}B_j,
\end{gather}
\end{subequations}
\normalsize
where we have defined $A_{j}=v_{j}^{+}+v_{j}^{-}$ and $B_{j}=v_{j}^{-}-v_{j}^{+}$ and where the explicit dependence on $k$ of the components $v_{\ell}^{\pm}(k)$ has been dropped for notational convenience. The recurrence system \eqref{Recu-Syst-5} is ''reversed''. The terminating conditions are at $\ell=k$. In this case, \eqref{Recu-Syst-5} becomes
\begin{gather*}
[2k+1+\zeta+i\xi]v_{k}^{+}=[\xi-i(2k+\zeta+1)]v_{k}^{-},\\
[2k+1+\zeta-i\xi]v_{k}^{-}=[\xi+i(2k+\zeta+1)]v_{k}^{+}.
\end{gather*}
Choosing $v_{k}^{-}=1$, we obtain the terminating conditions
\begin{align*}
v_{k}^{+}=-i,\qquad v_{k}^{-}=1.
\end{align*}
Upon taking
\begin{align*}
A_{\ell}=\alpha_0\widehat{A}_{\ell},\quad B_{\ell}=\beta_0\widehat{B}_{\ell},
\end{align*}
where $\alpha_0=(1-i)$ and $\beta_0=(1+i)$, it is easily seen that $\widehat{A}_{\ell}$ and $\widehat{B}_{\ell}$ are real and that the system \eqref{Recu-Syst-5} is equivalent to
\small
\begin{gather*}
[k+\mu_y+1/2]\widehat{A}_{\ell}=(\ell+\mu_y+1/2)\widehat{B}_{\ell}+2\mu_y\sum_{j=\ell+1}^{k}\widehat{B}_{j},\\
[k+\mu_x+1/2]\widehat{B}_{\ell}=(\ell+\mu_x+1/2)\widehat{A}_{\ell}+2\mu_x\sum_{j=\ell+1}^{k}(-1)^{j-\ell}\widehat{A}_{j},
\end{gather*}
\normalsize
with the terminating conditions $\widehat{A}_{k}=1$ and $\widehat{B}_{k}=1$. Introducing the reversed components $\widetilde{a}_{n}=\widehat{A}_{k-n}$, $\widetilde{b}_{n}=\widehat{B}_{k-n}$, we obtain the system
\begin{subequations}
\label{Recu-Syst-6}
\begin{align}
[k+\mu_y+1/2]\widetilde{a}_{n}&=[k-n+\mu_y+1/2]\widetilde{b}_{n}+2\mu_y\sum_{j=0}^{n-1}\widetilde{b}_{j},\\
[k+\mu_x+1/2]\widetilde{b}_{n}&=[k-n+\mu_x+1/2]\widetilde{a}_{n}+2\mu_x(-1)^{n}\sum_{j=0}^{n-1}(-1)^{j}\widetilde{a}_{j},
\end{align}
\end{subequations}
with the initial conditions $\widetilde{a}_0=1$ and $\widetilde{b}_0=1$. Taking into account all the preceding transformations, we have
\small
\begin{align}
\label{V-Odd}
v_{\ell}^{-}(k)=\frac{\alpha_0 \widetilde{a}_{k-\ell}+\beta_0\widetilde{b}_{k-\ell}}{2},\quad v_{\ell}^{+}(k)=\frac{\alpha_0\widetilde{a}_{k-\ell}-\beta_0\widetilde{b}_{k-\ell}}{2},
\end{align}
\normalsize
where $\widetilde{a}_{\ell}$, $\widetilde{b}_{\ell}$ are the unique solutions to the recurrence system \eqref{Recu-Syst-6}.
\subsubsection{The $\nu^{+}_k$ eigenvalue sector}\hfill\bigskip

\noindent
Let us now consider the eigenvectors $\ket{k,+}_{\mathcal{Q}}$ corresponding to the eigenvalue $\nu_{k}^{+}$ with expansion
\begin{align*}
\ket{k,+}_{\mathcal{Q}}=\sum_{\substack{\ell=0\\\sigma=\pm}}^{k}u_{\ell}^{\sigma}(k)\ket{\ell,\sigma},
\end{align*}
in the circular basis. Proceeding along the same lines as in the previous computation, we find that the terminating condition are of the form
\begin{align*}
v_{k}^{+}=\frac{i(2k+1+\zeta+i\xi)}{2k+1+\zeta-i\xi},\qquad v_{k}^{-}=1,
\end{align*}
and that the components are given by
\small
\begin{align}
\label{U-Odd}
u_{\ell}^{-}(k)=\frac{\gamma_0 \widetilde{a}_{k-\ell}+\epsilon_0\widetilde{b}_{k-\ell}}{2},\quad u_{\ell}^{+}(k)=\frac{\gamma_0\widetilde{a}_{k-\ell}-\epsilon_0\widetilde{b}_{k-\ell}}{2},
\end{align}
\normalsize
where
\begin{align*}
\gamma_0=\frac{(1+i)(2k+1+2\mu_y)}{2k+1+\zeta-i\xi},\quad \epsilon_0=\frac{(1-i)(2k+1+2\mu_x)}{2k+1+\zeta-i\xi} .
\end{align*}
\subsection{Generating functions and Heun polynomials}
As is seen from the formulas \eqref{V-Odd} and \eqref{U-Odd}, the main part of the components of the eigenvectors of $\mathcal{Q}$ in the circular basis is given by the solutions to the recurrence system \eqref{Recu-Syst-6}. As in section 4 for the $N$ even case, we bring the ordinary generating functions
\begin{align*}
\widetilde{A}(z)=\sum_{n}\widetilde{a}_{n}z^{n},\qquad \widetilde{B}(z)=\sum_{n}\widetilde{b}_{n}z^{n}.
\end{align*}
Upon using the identities \eqref{GenFun-Iden}, we obtain from \eqref{Recu-Syst-6} the associated differential system
\small
\begin{subequations}
\label{Diff-Syst-Odd}
\begin{align}
(k+\mu_y+1/2)\widetilde{A}(z)=(k-\mu_y+1/2-z\pd_{z})\widetilde{B}(z)+\frac{2\mu_y}{1-z}\widetilde{B}(z),\\
(k+\mu_x+1/2)\widetilde{B}(z)=(k-\mu_x+1/2-z\pd_{z})\widetilde{A}(z)+\frac{2\mu_x}{1+z}\widetilde{A}(z).
\end{align}
\end{subequations}
\normalsize
By direct substitution, we find that the generating functions are expressed in terms of the Heun functions
\begin{subequations}
\label{Heun-Odd}
\begin{align}
\label{Heun-Odd-a}
\widetilde{A}(z)&=(1+z)H\ell(a,q_A;\alpha_A,\beta_{A},\gamma_{A},\delta_{A},z),\\
\label{Heun-Odd-b}
\widetilde{B}(z)&=(1-z)H\ell(a,q_{B};\alpha_{B},\beta_{B},\gamma_{B},\delta_{B},-z),
\end{align}
\end{subequations}
where
\begin{subequations}
\label{Parameters-A}
\begin{gather}
a=-1,\quad q_{A}=2k(\mu_y-\mu_x-1),\quad \alpha_{A}=-2k,\\
\beta_{A}=\mu_x+\mu_y+1,\quad \gamma_{A}=-2k-\mu_x-\mu_y,\quad \delta_{A}=2\mu_y.
\end{gather}
\end{subequations}
and where the parameters $q_{B},\cdots \delta_{B}$ are obtained from \eqref{Parameters-A} by the transformation $\mu_x\leftrightarrow \mu_y$. The form of the parameters involved in the Heun functions show that once again one has a truncation at degree $2k+1$ and hence the Heun functions appearing in \eqref{Heun-Odd} are in fact Heun polynomials. The generating functions $\widetilde{A}(z)$ and $\widetilde{B}(z)$ are polynomials of degree $2k+1$.
\subsection{Expansion of Heun polynomials in complementary Bannai-Ito polynomials}
The expansion of the Heun polynomials can be obtained using the associated three-term recurrence relation. Since the expansion coefficients of $\widetilde{B}(z)$ and $\widetilde{A}(z)$ are related by the simple relation $z\leftrightarrow -z$ and $\mu_x\leftrightarrow \mu_y$, we shall focus on the expansion of $\widetilde{A}(z)$.

We examine the expansion of the Heun polynomial appearing in \eqref{Heun-Odd-a}
\begin{align}
\label{Dompe-4}
H\ell(a,q_{A},\alpha_A,\beta_{A},\gamma_{A},\delta_{A},z)=\sum_{r}\mathcal{P}_{r}z^{r},
\end{align}
where the parameters are given in \eqref{Parameters-A}.
Using the recurrence coefficients given in \eqref{Recurrence-Coeff}, one finds that the expansion coefficients $\mathcal{P}_{n}$ are polynomials of degree $r$ in $\xi=\mu_x-\mu_y$ that obey the recurrence relation
\begin{align}
\label{Dompe-3}
\sigma_{r+1}\mathcal{P}_{r+1}(\xi)+\kappa_{r}\mathcal{P}_{r-1}(\xi)=(1+\xi)\mathcal{P}_{r}(\xi),
\end{align}
with $\mathcal{P}_{-1}=0$, $\mathcal{P}_{0}=1$ and where
\begin{align*}
\sigma_{r+1}=\frac{(r+1)(N+\zeta-r)}{2r-N},\quad \kappa_{r}=\frac{(N+1-r)(r+\zeta)}{2r-N},
\end{align*}
where we have taken $N=2k$. Introducing the monic polynomials $\mathcal{P}_{r}(\xi)=\frac{\widehat{\mathcal{P}}_{r}(\xi)}{\sigma_{1}\cdots \sigma_{r}}$, we obtain
\begin{align}
\label{Recu-2}
\widehat{\mathcal{P}}_{r+1}(\xi)+u_{r}\widehat{\mathcal{P}}_{r-1}(\xi)=(1+\xi)\widehat{\mathcal{P}}_{r}(\xi)
\end{align}
where 
\begin{align*}
u_{r}=-\frac{r(N+1-r)(N+\zeta-r-1)(r+\zeta)}{(N-2r)(N-2r+2)}.
\end{align*}
Upon comparing \eqref{Recu-2} with the recurrence coefficients for the CBI polynomials given in \eqref{CBI-Coeff}, it is directly seen that the polynomials defined by the recurrence \eqref{Recu-2} correspond to CBI polynomials with the parametrization
\begin{align}
\label{Dompe-2}
\rho_1=\frac{\zeta-1}{2},\quad r_1=\frac{2k+\zeta+1}{2},\quad \rho=0,\quad r_2=0.
\end{align}
Hence, when $r\leqslant k$, we have
\begin{align}
\label{KC-1}
\mathcal{P}_{r}(\xi)=\frac{(-1)^{r}4^{r}}{r!}\frac{(k+1-r)_{r}}{(2k+\zeta+1-r)_{r}}I_{r}((1+\xi)/2;\rho_1,\rho_2,r_1,r_2),\quad r\leqslant k,
\end{align}
where $\rho_1$, $\rho_2$, $r_1$ and $r_2$ are given by \eqref{Dompe-2} and $I_{n}(x;\rho_1,\rho_2,r_1,r_2)$ are the complementary Bannai-Ito polynomials. As can be seen by the recurrence relation \eqref{Dompe-3}, the expansion coefficients of the Heun function \eqref{Dompe-4} truncate at order $2k+1$ and are center-symmetric. Hence, for $r>k$, we have
\begin{align}
\label{KC-2}
\mathcal{P}_{r}(\xi)=\mathcal{P}_{2k+1-r}(\xi),\qquad r>k.
\end{align}
Taking into account the relation between $\widetilde{A}(z)$ and $\widetilde{B}(z)$, the expansion coefficients of the Heun function appearing in \eqref{Heun-Odd-b}, denoted by $\mathcal{T}_{n}(\xi)$, are easily seen to be
\begin{gather}
\label{KC-3}
\mathcal{T}_{r}(\xi)=\frac{4^{r}}{r!}\frac{(k+1-r)_{r}}{(2k+\zeta+r-1)_{r}}I_{r}((1-\xi)/2;\rho_1,\rho_2,r_1,r_2),\quad r\leqslant 2k,\\
\label{KC-4}
\mathcal{T}_{r}(\xi)=\mathcal{T}_{2k+1-r}(\xi),\quad r>2k.
\end{gather}
Collecting all the previous results, we write
\begin{subequations}
\label{Final-Recu}
\begin{gather}
\widetilde{a}_{n}=\mathcal{P}_{n}(\xi)+\mathcal{P}_{n-1}(\xi),\\
\widetilde{b}_{n}=\mathcal{T}_{n}(\xi)+\mathcal{T}_{n-1}(\xi),
\end{gather}
\end{subequations}
where $\mathcal{P}_{-1}=0$, $\mathcal{T}_{-1}=0$ and $\mathcal{P}_{n}(\xi)$, $\mathcal{T}_n(\xi)$ are given by \eqref{KC-1}, \eqref{KC-2}, \eqref{KC-3} and \eqref{KC-4}.
\subsection{Eigenvectors of $J_2$}
To obtain the expansion of the eigenvectors of $J_2$ in the circular basis, one must relate the eigenvectors of the operator $\mathcal{Q}$ to the eigenvectors of $J_2$. This relation has been obtained in the previous paper \cite{Genest-2012-4}. We have
\begin{align*}
\ket{k,\pm}_{\mathcal{Q}}=\frac{1}{\sqrt{2}}\left(\ket{k,+}_{J_2}\mp\upsilon_{k}\ket{k,-}_{J_2}\right)
\end{align*}
where $\ket{k,\pm}_{J_2}$ are the eigenvectors of $J_2$ corresponding to the eigenvalues
\begin{align*}
\lambda_{k}^{\pm}=\pm\sqrt{(k+\mu_x+1/2)(k+\mu_y+1/2)},\qquad k=0,\ldots,m,
\end{align*}
and where
\small
\begin{align*}
v_{k}=\left[\frac{\xi+2i\sqrt{(k+\mu_x+1/2)(k+\mu_y+1/2)}}{2k+\zeta+1}\right].
\end{align*}
\normalsize
The inverse relation reads
\begin{subequations}
\label{Relation-J2}
\begin{align}
\ket{k,+}_{J_2}=\frac{1}{\sqrt{2}}\left(\ket{k,+}_{\mathcal{Q}}+\ket{k,-}_{\mathcal{Q}}\right),\\
\ket{k,-}_{J_2}=\frac{-1}{v_{k}\sqrt{2}}\left(\ket{k,+}_{\mathcal{Q}}-\ket{k,-}_{\mathcal{Q}}\right)
\end{align}
\end{subequations}
Using the relations \eqref{Relation-J2}, the results \eqref{Final-Recu}, \eqref{U-Odd} and \eqref{V-Odd}, one has an explicit expression for the expansion of the eigenvectors of $J_2$ in the circular basis for the case $N$ odd.
\subsection{The fully isotropic case : $-1$ Jacobi polynomials}
We consider again the case $\mu_x=\mu_y=\mu$ which corresponds to the fully isotropic Dunkl oscillator, where two independent identical parabosonic oscillators are combined. We return to the system of differential equations for the generating functions of the components of the eigenvectors of $\mathcal{Q}$ given in \eqref{Diff-Syst-Odd}. When $\mu_x=\mu_y=\mu$, one has
\begin{gather*}
(k+\mu+1/2)\widetilde{A}(z)=(k-\mu+1/2-z\pd_{z})\widetilde{B}(z)+\frac{2\mu}{1-z}\widetilde{B}(z),\\
(k+\mu+1/2)\widetilde{B}(z)=(k-\mu+1/2-z\pd_{z})\widetilde{A}(z)+\frac{2\mu}{1+z}\widetilde{A}(z).
\end{gather*}
It is easily seen from the above formulas that $\widetilde{A}(z)=\widetilde{B}(-z)$. Hence the generating function $\widetilde{A}(z)$ satisfies the differential equation
\begin{align*}
(k+\mu+1/2)\widetilde{A}(z)=(k-\mu+1/2-z\pd_{z})\widetilde{A}(-z)+\frac{2\mu}{1-z}\widetilde{A}(-z),
\end{align*}
which may be cast in the form of an eigenvalue equation
\begin{align*}
L\widetilde{A}(z)=4\mu \widetilde{A}(z),
\end{align*}
where
\begin{align*}
L=2(1-z)\pd_{z}R+\left[(-2k-1+2\mu)+\frac{2k+1+2\mu}{z}\right](\mathbb{I}-R),
\end{align*}
where $Rf(z)=f(-z)$. It is recognized that the operator $L$ is a special case of the defining operator of the little $-1$ Jacobi polynomials \cite{Vinet-2011}. 

The little $-1$ Jacobi polynomials, denoted by $P_{n}^{-1}(x)$, obey the eigenvalue equation
\begin{align*}
\Omega P_{n}(x)=\lambda_{n}P_{n}(x),\qquad \lambda_{n}=
\begin{cases}
-2n & \text{$n$ even}\\
2(\alpha+\beta+n+1) & \text{$n$ odd}                                           
\end{cases},
\end{align*}
where
\begin{align*}
\Omega=2(1-x)\pd_{x}R+(\alpha+\beta+1-\alpha/x)(\mathbb{I}-R).
\end{align*}
Comparing the operators $\Omega$ and $L$, it is seen that the generating function $\widetilde{A}(z)$ corresponds to a $-1$ Jacobi polynomial of degree $n=2k+1$ with parameters
\begin{align*}
\alpha=-2k-2\mu-1,\qquad \beta=4\mu-1.
\end{align*}
Using this identification and the explicit formula for the little $-1$ Jacobi polynomials derived in \cite{Vinet-2011}, we obtain
\begin{align*}
\widetilde{A}(z)=
\pFq{2}{1}{-k,\mu}{-\mu-k}{z^2}+\frac{\mu z}{k+\mu}\,\pFq{2}{1}{-k,\mu+1}{-k-\mu+1}{z^2}.
\end{align*}
This directly yields the following result for the recurrence coefficients $\widetilde{a}_{n}$:
\begin{align*}
\widetilde{a}_{2n}=\frac{(-k)_{n}(\mu)_{n}}{n!(-\mu-k)_{n}},\quad \widetilde{a}_{2n+1}=\frac{\mu}{k+\mu}\frac{(-k)_{n}(\mu+1)_{n}}{n!(-\mu-k+1)_{n}}.
\end{align*}
Using the symmetry $\widetilde{B}(z)=\widetilde{A}(-z)$, we also obtain
\begin{align*}
\widetilde{b}_{2n}=\frac{(-k)_{n}(\mu)_{n}}{n!(-\mu-k)_{n}},\quad \widetilde{b}_{2n+1}=\frac{-\mu}{k+\mu}\frac{(-k)_{n}(\mu+1)_{n}}{n!(-\mu-k+1)_{n}}.
\end{align*}
Once again, the components of the eigenvectors of $\mathcal{Q}$ drastically simplify in the isotropic case $\mu_x=\mu_y=\mu$. Moreover, the preceding computations entail a relation between a special case of Heun polynomials and the little $-1$ Jacobi polynomials.

\section{Representations of $sd(2)$ in the $J_2$ eigenbasis}
We now investigate the representation space in which the operator $J_2$ is diagonal. The matrix elements of the generators of the Schwinger-Dunkl algebra will be derived using the defining relations of $sd(2)$, which read
\begin{gather*}
\{J_1,R_{x_i}\}=0,\quad \{J_2,R_{x_i}\}=0,\quad [J_3,R_{x_i}]=0,\\
\label{Commutation-2}
[J_2,J_3]=iJ_1,\qquad [J_3,J_1]=iJ_2,\\
\label{Commutation-3}
[J_1,J_2]=i[J_3+J_3(\mu_x R_{x}+\mu_y R_y)-\mathcal{H}(\mu_xR_x-\mu_y R_y)/2],
\end{gather*}
where $R_x^2=R_y^2=\mathbb{I}$. It will prove convenient to treat the even and odd dimensional representations separately.
\subsection{The $N$ odd case}
We first consider the case where $N$ is odd. The representation space $\mathcal{C}$ is spanned in this case by the basis vectors $\ket{k,\pm}$ with $k\in\{0,\ldots,m\}$ on which the generator $J_2$ acts in a diagonal fashion
\begin{align*}
J_2\ket{k,\pm}=\lambda_{k}^{\pm}\ket{k,\pm},\qquad k=0,\ldots,m,
\end{align*}
where $m=(N-1)/2$ and where the eigenvalues of $J_2$, derived in section 3, are given by
\begin{align*}
\lambda_{k}^{\pm}=\pm \sqrt{(k+\mu_x+1/2)(k+\mu_y+1/2)},\quad k=0,\ldots,m.
\end{align*}
Since $R_{x}$, $R_{y}$ anti-commute with $J_2$ and given that $R_{x}R_{y}$ is central in the algebra $sd(2)$, we can take
\begin{align*}
R_{x}\ket{k,\pm}=\epsilon \ket{k,\mp},\qquad R_{y}\ket{k,\pm}=\ket{k,\mp},
\end{align*}
where $\epsilon=\pm 1$. Here we choose $\epsilon=-1$, which corresponds to the representation encountered in the model. In the basis $\{\ket{0,+},\ket{0,-},\cdots,\ket{m,+},\ket{m,-}\}$, the matrices representing the involutions $R_{x}$, $R_{y}$ have the form
\begin{align*}
R_y=-R_{x}=\mathrm{diag}(\sigma_1,\ldots,\sigma_1),\qquad \sigma_1=
\begin{pmatrix}
0 & 1\\
1 & 0
\end{pmatrix},
\end{align*}
which is identical to their action in the circular basis. The Hamiltonian $\mathcal{H}$ has the action
\begin{align*}
\mathcal{H}\ket{k,\pm}=(N+\mu_x+\mu_y+1)\ket{k,\pm}.
\end{align*}
The action of the operator $J_3$ on this representation space can be derived by imposing the commutation relation \eqref{Commutation-3} using \eqref{Commutation-2} to define $J_1$. The action of $J_3$ on the basis $\ket{k,\pm}$ is taken to be
\begin{align*}
J_3\ket{k,+}=\sum_{j=0}^{m}M_{jk}^{\pm}\ket{j,\pm},\quad J_3\ket{k,-}=\sum_{j}N_{jk}^{\pm}\ket{j,\pm}.
\end{align*}
With these definitions, it is easily seen that the commutation relation \eqref{Commutation-3} is equivalent to the following system of relations
\small
\begin{subequations}
\label{Coupled}
\begin{align}
\label{a}
[(\lambda_{k}^{+})^2-2(\lambda_{j}^{+}\lambda_{k}^{+})+(\lambda_{j}^{+})^2-1]M_{jk}^{+}&=(\mu_y-\mu_x)N_{jk}^{+},\\
\label{b}
[(\lambda_{k}^{+})^2-2(\lambda_{j}^{-}\lambda_{k}^{+})+(\lambda_{j}^{-})^2-1]M_{jk}^{-}&=(\mu_y-\mu_x)N_{jk}^{-}+\delta_{jk}\frac{\zeta(N+1+\zeta)}{2},\\
\label{c}
[(\lambda_{k}^{-})^2-2(\lambda_{j}^{+}\lambda_{k}^{-})+(\lambda_{j}^{+})^2-1]N_{jk}^{+}&=(\mu_y-\mu_x)M_{jk}^{+}+\delta_{jk}\frac{\zeta(N+1+\zeta)}{2},\\
\label{d}
[(\lambda_{k}^{-})^2-2(\lambda_{j}^{-}\lambda_{k}^{-})+(\lambda_{j}^{-})^2-1]N_{jk}^{-}&=(\mu_y-\mu_x)M_{jk}^{-},
\end{align}
\end{subequations}
\normalsize
where the relations \eqref{a},\eqref{b} were obtained by acting on $\ket{k,+}$ and the relations \eqref{c},\eqref{d} by acting on $\ket{k,-}$. It follows directly from the solution of the system \eqref{Coupled} that $J_3$ acts in a six-diagonal fashion on the eigenbasis of $J_2$. For $j=k$, we obtain
\small
\begin{subequations}
\label{central}
\begin{gather}
M_{kk}^{+}=\frac{\xi\zeta(N+\zeta+1)}{2(2k+\zeta)(2k+\zeta+2)},\quad N_{kk}^{+}=\frac{\zeta(N+\zeta+1)}{2(2k+\zeta)(2k+\zeta+2)},\\
M_{kk}^{-}=\frac{\zeta(N+\zeta+1)}{2(2k+\zeta)(2k+\zeta+2)},\quad N_{kk}^{-}=\frac{\xi\zeta(N+\zeta+1)}{2(2k+\zeta)(2k+\zeta+2)},
\end{gather}
\end{subequations}
\normalsize
where $\xi=\mu_x-\mu_y$, $\zeta=\mu_x+\mu_y$. For $j=k+\ell$ or $j=k-\ell$ with $\ell>1$, only the trivial solution occurs, so the matrix representing $J_3$ in the eigenbasis of $J_2$ is block tridiagonal with all block $2\times$ 2. Hence it acts in a six-diagonal fashion on the eigenbasis of $J_2$. Using the commutation relations and the system \eqref{Coupled}, it is possible to obtain an expression for the matrix elements of $J_3$ which involves a set of arbitrary non-zero parameters  $\{\beta_{n}\}$ for $n=0,\ldots,m$. After considerable algebra, one finds that the matrix $J_3$ has the form
\small
\begin{align*}
J_3=
\begin{pmatrix}
C_{0}& U_{1} & & &\\
D_{0} & C_{1} & U_{2} & &\\
 & D_1 & C_{2}& \ddots &\\
 & &\ddots & \ddots & U_{m}\\
&&&D_{m-1}&C_{m} 
\end{pmatrix},
\end{align*}
\normalsize
where the blocks are given by
\small
\begin{align*}
U_{k}=
\beta_{k}\begin{pmatrix}
M_{k-1k}^{+} & 1 \\
1 & M_{k-1k}^{+} 
\end{pmatrix},
\quad
C_{k}=
\begin{pmatrix}
M_{kk}^{+} & N_{kk}^{+}\\
M_{kk}^{-} & N_{kk}^{-}
\end{pmatrix},
\quad
D_{k}=
\beta_{k+1}^{-1}
\begin{pmatrix}
M_{k+1k}^{+} & N_{k+1k}^{+}\\
N_{k+1k}^{+} & M_{k+1k}^{+} 
\end{pmatrix}.
\end{align*}
\normalsize
The matrix elements of the central blocks are given by \eqref{central}. The components of the upper blocks $U_{k}$ have the form
\small
\begin{gather*}
M_{k-1k}^{+}=\frac{1}{2\xi}\left[1-4(k+\mu_x)(k+\mu_y)-4\left\{(k+\mu_x-1/2)_{2}(k+\mu_y-1/2)_{2}\right\}^{1/2}\right].
\end{gather*}
\normalsize
The matrix elements of the lower blocks have the form
\small
\begin{subequations}
\begin{gather*}
M_{k+1k}^{+}=\frac{\xi(k+1)(2k-N+1)(k+1+\zeta)(2k+2\zeta+N+3)}{4(2k+\zeta+2)(2k+\zeta+1)_{3}},\\
N_{k+1k}^{+}=E_{k}M_{k+1k}^{+}.
\end{gather*}
\end{subequations}
\normalsize
where
\small
\begin{align*}
E_k=\frac{1}{2\xi}\left[1-4(k+\mu_x+1)(k+\mu_y+1)+4\{(k+\mu_x+1/2)_{2}(k+\mu_y+1/2)_2\}^{1/2}\right].
\end{align*}
\normalsize
We note that these matrix elements are valid for $\mu_x\neq \mu_y$. In the latter case, the form of the spectrum of $J_2$ changes and the computation has to be redone from the start.
\subsection{The $N$ even case}
We now consider the $N$ even case. The representation space $\mathcal{C}$ is spanned in this case by the basis vectors $\ket{0,-}$ and $\ket{k,\pm}$ with $k=1,\ldots,m$ on which the operator $J_2$ acts in a diagonal fashion
\begin{align*}
J_2\ket{k,\pm}=\lambda^{\pm}_{k}\ket{k,\pm},\qquad k=0,\ldots,m,
\end{align*}
where the eigenvalues of $J_2$, determined in section 3, are given by the formula
\begin{align*}
\lambda_{k}^{\pm}=\pm\sqrt{k(k+\zeta)},\quad k=0,\ldots,m.
\end{align*}
Note that the eigenvalue $\lambda_0$ is non-degenerate. In this representation, we choose the following action for the involutions $R_{x}$, $R_{y}$:
\begin{align*}
R_{x_i}\ket{0,-}=\ket{0,-},\quad R_{x_i}\ket{k,\pm}=\ket{k,\mp},
\end{align*}
and hence the reflections have the matrix representation
\begin{align*}
R_{x}=R_{y}=\mathrm{diag}(1,\sigma_1,\ldots,\sigma_1),
\quad
\sigma_1=
\begin{pmatrix}
0 & 1 \\
1 & 0
\end{pmatrix}.
\end{align*}
The central element (Hamiltonian) $\mathcal{H}$ has the familiar action
\begin{align*}
\mathcal{H}\ket{k,\pm}=(N+\mu_x+\mu_y+1)\ket{k,\pm}.
\end{align*}
Following the same steps as in $(6.1)$, a direct computation shows that the in this case $J_3$ has the matrix representation
\small
\begin{align*}
J_3=
\begin{pmatrix}
c_0& u_{1} & & & \\
d_{0}& C_{1} & U_{2}& &\\
 & D_{1} & C_{2} & \ddots & \\
& & \ddots & \ddots & U_{m} \\
&&&D_{m-1}&C_{m}
\end{pmatrix}.
\end{align*}
\normalsize
The special $2\times 1$ and $1\times 1$ blocks are given by
\small
\begin{gather*}
c_0=\frac{\xi(N+\zeta+1)}{2(1+\zeta)},\quad
u_1=
\begin{pmatrix}
\alpha_1 & \alpha_1
\end{pmatrix},
\quad
d_0=
\alpha_1^{-1}
\begin{pmatrix}
w_{N} & w_{N}
\end{pmatrix}^{t},
\end{gather*}
\normalsize
with
\small
\begin{gather*}
w_{N}=\frac{(N/2)(1+2\mu_x)(1+2\mu_y)(N/2+\zeta+1)}{2(1+\zeta)^2(2+\zeta)}.
\end{gather*}
\normalsize
The $2\times 2$ blocks have the form
\small
\begin{align*}
U_{k}=\alpha_{k}
\begin{pmatrix}
1 & N_{k-1k}^{+} \\
N_{k-1k}^{+} & 1
\end{pmatrix},
\quad
C_{k}=
\begin{pmatrix}
M_{kk}^{+} & N_{kk}^{+}\\
N_{kk}^{+} & M_{kk}^{+}
\end{pmatrix},
\quad
D_{k}=\alpha_{k+1}^{-1}
\begin{pmatrix}
M_{k+1k}^{+} & N_{k+1k}^{+}\\
N_{k+1k}^{+}& M_{k+1k}^{+}
\end{pmatrix}.
\end{align*}
\normalsize
where
\footnotesize
\begin{gather*}
M_{kk}^{+}=\frac{\xi\zeta (N+\zeta+1)}{2(2k-1+\zeta)(2k+1+\zeta)},\quad
N_{kk}^{+}=\frac{-\xi(N+\zeta+1)}{2(2k-1+\zeta)(2k+1+\zeta)},\\
N_{k-1k}^{+}=\zeta^{-1}\left\{\zeta+2(k-1)(k+\zeta)-2[(k-1)_{2}(k-1+\zeta)_2]^{1/2}\right\},\\
M_{k+1k}^{+}=\frac{(N/2-k)(N/2+k+1+\zeta)(2k+1+2\mu_x)(2k+1+2\mu_y)\left\{\zeta+2k(k+\zeta+1)+2[(k)_{2}(k+\zeta)_{2}]^{1/2}\right\}}{4(2k+1+\zeta)(2k+\zeta)_3},\\
N_{k+1k}^{+}=\frac{\zeta(N/2-k)(N/2+k+\zeta+1)(2k+1+2\mu_x)(2k+1+2\mu_y)}{4(2k+\zeta+1)(2k+\zeta)_{3}}.
\end{gather*}
\normalsize
The parameters of the sequence $\{\alpha_k\}$ are arbitrary but non-zero; they could be fixed, for example, by examining the action of $J_2$ on the eigenstates of the 2D Dunkl oscillator in the polar coordinate representation. We have thus obtained the action of the operator $J_3$ on the eigenstates of $J_2$. Recall that $J_3$ is the symmetry operator associated to the separation of variables in Cartesian coordinates and $J_2$ is the symmetry associated to the separation of variables in polar coordinates.
\section{Conclusion}
We have investigated the finite-dimensional irreducible representations of the Schwinger-Dunkl algebra $sd(2)$, which is the symmetry algebra of the two-dimensional Dunkl oscillator in the plane. The action of the symmetry generators in the representations were obtained in three different bases. In the Cartesian basis, the symmetry generator $J_3$ associated to separation of variables in Cartesian coordinates is diagonal, and the symmetry $J_2$ is tridiagonal. In the circular basis, the operator $J_3$ acts in a three-diagonal fashion and $J_2$ has a block upper-triangular structure with all blocks $2\times 2$. The eigenvalues of $J_2$ can be evaluated algebraically in the circular basis and the expansion coefficients for the eigenvectors of $J_2$ in this basis are generated by Heun polynomials and are expressed in terms of the para-Krawtchouk polynomials. Finally, it was shown that in the eigenbasis of $J_2$, the operator $J_3$ acts in a block tridiagonal fashion with all blocks $2\times 2$, that is, that $J_3$ is six-diagonal.

It has been seen that the Dunkl oscillator model is superintegrable and closely related to the $-1$ orthogonal polynomials of the Bannai-Ito scheme. In this connection, the study of the 3D Dunkl oscillator model and the singular 2D Dunkl oscillator could also provide additional insight in the physical interpretation of the orthogonal polynomials of the Bannai-Ito scheme.

\end{document}